%% file: eymd.tex
\documentclass{article}
\usepackage{amsmath,amsthm,amssymb,array,latexsym,cite,epsfig,finalT}

\numberwithin{equation}{section}

\newcommand{\sauthor}{T.A. Oliynyk}
\newcommand{\stitle}{Newtonian perturbations and the EYMd equations}

\input eymd.def

\begin{document}


\input eymd_tit.tex
\input intro.tex
\input feqns.tex

\input wsob.tex

\input sss.tex
\input newt.tex

\input smooth.tex

\input reduced.tex

\input ymd.tex

\input eymdsol.tex

\input conc.tex

\bigskip

\noindent \emph{Acknowledgments}.  I would like to thank R. Bartnik
for helpful discussions and advice. I would also like to
thank the referees for their useful criticisms and comments.
This work was partially supported
by the ARC grant A00105048 at the University of Canberra and
by the NSERC grants A8059 and 203614 at the University of Alberta.


\input eymd.bbl
\end{document}

%% file: eymd_tit.tex
\title{Newtonian perturbations and the
Einstein-Yang-Mills-dilaton equations
}
\author{
Todd A. Oliynyk \thanks{Present address: School of Mathematical Sciences, Monash University, VIC 3800 Australia}
\thanks{todd.oliynyk@sci.monash.edu.au} \\
Department of Mathematical and Statistical Sciences\\
University of Alberta\\ Edmonton AB T6G 2G1}
\date{}
\maketitle
\begin{abstract}
In this paper we show that problem of
proving the existence of a countable number of
solutions to the static spherically symmetric $SU(2)$
Einstein-Yang-Mills-dilaton (EYMd) equations can be reduced to proving the non-existence of
solutions to the linearized Yang-Mills-dilaton equations (lYMd)
satisfying certain asymptotic conditions.
The reduction from a non-linear to a linear problem is
achieved using a Newtonian perturbation type argument.
\end{abstract}

%% file: intro.tex
\sect{intro}{Introduction}

Unlike the four dimensional Yang-Mills (YM) equations which have
no static solutions of finite energy \cite{Dess,Cole}, the 
Euclidean $SU(2)$ Yang-Mills-dilaton (YMd) equations were shown numerically to 
possess a countably infinite sequence
of static, globally regular, spherically symmetric solutions 
\cite{LavMai92,Bizo93a}. Existence of these solutions
was rigorously established using shooting techniques in
\cite{HMT95}. 
The dilaton play the role of an attractive force which counterbalances
the repulsive nature of the Yang-Mills fields and this makes it possible
for static solutions to exist on flat space. This is a simpler
situation compared to the more well known  Bartnik-McKinnon (BK) static solutions 
in Einstein-Yang-Mills (EYM) theory \cite{BK} 
where gravity is the counterbalancing force. 

In the papers \cite{LavMai93,Bizo93b} it was found, again numerically, that the 
YMd solutions
persist when the Yang-Mills and dilaton fields are
coupled to gravity. The result is a countably infinite sequence
of static, globally regular, spherically symmetric solutions
to the $SU(2)$ Einstein-Yang-Mills-dilaton (EYMd) equations with the
same qualitative behavior for the Yang-Mills and dilaton fields
as when gravity is absent. These solutions limit to the BK solutions
as the dilaton coupling constant goes to zero which helps to
explain why the Yang-Mills fields for the EYMd solutions have
a similar behavior to the BK solutions where there is no dilaton field.

In this paper we show that problem of 
proving the existence of a countably infinite number of
solutions to the static spherically symmetric $SU(2)$
EYMd equations can be reduced to proving the non-existence of
solutions to the linearized Yang-Mills-dilaton equations (lYMd)
satisfying certain asymptotic conditions.
The reduction from a non-linear to a linear problem is 
achieved using a \emph{Newtonian perturbation} type argument.
Unfortunately, we have not been able exclude the possibility that
there exists solutions to the lYMd
that satisfy the asymptotic conditions. The main reason for
this is that the YMd solutions obtained
in \cite{HMT95} about which we
linearize are
unstable due to the presence of a negative
part of the spectrum for the lYMd operator. This means
that one cannot expect that a simple integration by parts
argument will work to rule out solutions to the linearized equations.
Instead we have to directly analyze the linearized
equations 
which is difficult because we
do not have much information about the YMd solutions
other than
that they exist and some asymptotic behavior.
However, we 
conjecture that there does not exist solutions
to the lYMd equations that satisfy the required asymptotic
conditions.
If this were the case then we would have a full existence proof.

Although the static spherically symmetric solutions to
the $SU(2)$-EYMd equations are unstable,
they may still be physically relevant as stringy 
generalizations of the BK solutions
which are very similar to sphalerons \cite{GalVol91}. Indeed,
sphalerons, which are unstable static solutions of the classical
equations for the bosonic sector of the electroweak theory,
are believed to be responsible for
violations of the conservation of baryon numbers at high
temperatures \cite{Ring88,ArnMcL87}. Therefore, it is possible
that the static EYMd solutions could  play a role in the
violation of the conservation of baryon
and lepton numbers at high temperatures.

The Newtonian perturbation argument in the form that is employed
in this paper was developed by Lottermoser in \cite{Lott92}
and subsequently used by Heilig to establish the existence
of slowly rotating stars \cite{Heil95}. It was also used
by the author to provide an existence proof for the
gravitating BPS monopole \cite{Oli2003}. These results
and the results of this paper show that Newtonian perturbation
method
is a useful approach to take in investigating
the  existence problem in general relativity for static
or stationary matter models. In addition to establishing existence,
the method also provides an analytic deformation from
a Newtonian solution to its general relativistic counterpart.
The deformation parameter can be interpreted as $1/c^{2}$ where
$c$ is the speed of light. So a Taylor expansion  in  $1/c^2$ can
be considered as a converging post-Newtonian expansion.
In this way, the Newtonian perturbation argument can be thought of
as the inverse of the Newtonian limit were Newtonian solutions
are obtained from general relativistic ones via the limit
$1/c^{2} \rightarrow 0$.  An attractive feature of the method is that
it produces solutions
to the Einstein field equations where the
matter fields are uniformly close to the their corresponding
Newtonian ones. This means that the properties of the Newtonian
solution pass directly to the corresponding relativistic solution.

The approach we take to establishing existence of static spherically
symmetric solutions is different from previous approaches
which rely on the fact that in spherical symmetry the static
Einstein equations reduce to ordinary differential equations
to which dynamical systems theory can be applied. The papers
\cite{SW,SWYM,BFM} which contain
existence proofs for the BK solutions in EYM theory exemplify
the dynamical systems approach to existence. The main advantage
of our approach is that it is in principle not restricted to spherical
symmetry which is important considering that it is known, numerically at least, 
that static axially symmetric solutions to the EYMd
equations exist \cite{KK98}. We note that 
a large amount of work has been done on gravitating gauge fields
both numerically and analytically starting with the pioneering 
work of Bartnik and McKinnon. For a comprehensive review see
\cite{VG}

This paper is organized as follows:
in section \ref{feqns} we set up the equations in a form
suitable to use the Newtonian perturbation method
while in section \ref{wsob}
we review the theory of weighted Sobolev spaces.
The Banach spaces for our field
variables (i.e. the dilaton field, gauge potential, and metric density)
are set up in section \ref{sss} and then in section \ref{smooth} the
field equations are shown to be smooth on those spaces. In section
\ref{newt} we discuss the YMd solutions of \cite{HMT95} and
their asymptotic properties.
Sections 
\ref{redsol}-\ref{eymdsol} contain the Newtonian perturbation argument.
In these sections it is shown that if there are no
solutions to the lYMd equations satisfying certain asymptotic
conditions then the static spherically symmetric YMd solutions of \cite{HMT95}
can be continued smoothly to static spherically symmetric 
solutions of the full EYMd equations.

%% file: feqns.tex
\sect{feqns}{EYMd equations}

For indexing of tensors and related quantities
Greek indices, $\alpha,\beta,\gamma$ etc., will always
run from $0$ to $4$ while Roman indices, $i,j,k$ etc.,
will range from $1$ to $3$. We will use bold letters such
as $\xb$ to denote points in $\Rbb^{3}$, i.e. $\xb=(x^{1},x^{2},x^{3})$.

Let $\underset{o}{g}$ denote the Minkowski metric on $\Rbb^{4}$.
Fix a global coordinate system $(x^{0},x^{1},
x^{2},x^{3})$ so that
\leqn{gflatup}{
\underset{o}{g}{}_{\alpha \beta} = \diag(-\lambda^{-1},1,1,1)
}
where $\lambda$ is a dimensionless parameter. From the way $\lambda$
appears in the metric \eqref{gflatup} it is useful 
to regard $\lambda$ as acting
like $1/c^2$ in which case the limit $\lambda \rightarrow 0$
can be thought of as an analogue of the \emph{Newtonian limit}.  
Define $\underset{o}{g}^{\alpha \beta}$
by $(\underset{o}{g}^{\alpha \beta}) := (\underset{o}{g}{}_{\alpha \beta})^{-1}$
which gives   
\leqn{gflatdwn}{
\underset{o}{g}{}^{\alpha \beta} = \diag(-\lambda,1,1,1) \, .
}
Define the Minkowski metric density
\leqn{godens}{
\go^{\alpha\beta} :=|\underset{o}{g}|^{\frac{1}{2}} 
\underset{o}{g}{}^{\alpha \beta} 
\quad \text{where} \quad |\underset{o}{g}| := 
|\det(\underset{o}{g}{}_{\alpha \beta})| \, .
}

Assume that $g_{\alpha \beta}$ is another metric defined
on $\Rbb^{4}$. Let $(g^{\alpha \beta}):= (g_{\alpha \beta})^{-1}$ and
introduce the density
\leqn{gdens}{
\g^{\alpha \beta} := |g|^{\frac{1}{2}} g^{\alpha\beta} 
\quad \text{where} \quad |g| := |\det(g_{\alpha \beta})| \, .
}
Following Lottermoser \cite{Lott92}, we form the tensor density
\leqn{udens}{
\U^{\alpha\beta} := \frac{1}{4\lambda^{\frac{3}{2}}}(\g^{\alpha \beta}
-\go^{\alpha \beta} ) \, 
}
which will be taken as our primary gravitational variable.
Observe that the metric
$g^{\alpha\beta}$ can be recovered from $\U^{\alpha\beta}$ by
\eqn{metfromdens1}{
g^{\alpha \beta} = \frac{1}{\sqrt{|g|}}\g^{\alpha\beta}
}
where
$\g^{\alpha \beta} = \go^{\alpha\beta} + 4\lambda^{\frac{3}{2}}\U^{\alpha\beta}$
and $|g| = |\det(\g^{\alpha\beta})|$.

Letting $G$ be a fixed constant and $\lambda^{2} G$ 
be the gravitational coupling constant, the Einstein equations 
\leqn{einstA}{
G_{\alpha\beta} = 8\pi \lambda^2 G T_{\alpha\beta}
}
can be written in terms
of the density \eqref{udens} as \cite{Lott92},
\leqn{einst}{
4\pi G |\df| T^{\alpha \beta} = A^{\alpha\beta}
+ B^{\alpha \beta} + C^{\alpha\beta} + D^{\alpha \beta} \, ,
}
where
\lalign{einsta}{
\gob^{\alpha \beta} & := \sqrt{\lambda} \go^{\alpha \beta}   , \label{einsta1} \\
\gob{}_{\alpha \beta} & := \sqrt{\lambda} \go{}_{\alpha \beta} \quad \text{where}
\quad
(\go{}_{\alpha \beta}) := (\go^{\alpha \beta})^{-1}  ,\label{einsta2} \\
\gb^{\alpha \beta} & := \sqrt{\lambda} \g^{\alpha \beta} =
\gob^{\alpha \beta} + 4\lambda^{2} \U^{\alpha \beta}  , \label{einsta3}\\
\gb_{\alpha \beta} & := \sqrt{\lambda} \g_{\alpha \beta} \quad \text{where} \quad
(\g_{\alpha\beta}) := (\g^{\alpha \beta})^{-1} \, , \label{einsta4} \\
\df &:= \lambda \det(\g^{\alpha\beta})  ,\label{einsta5} \\
A^{\alpha \beta}& := 2\left(\Half \gb_{\mu\nu}
\gb_{\gamma\rho} - \gb_{\rho\mu}\gb_{\gamma\nu}\right)\left(
\gb^{\alpha\kappa}\gb^{\beta\sigma}- \Half \g^{\alpha \beta}\gb^{\kappa\sigma}\right)
\U^{\mu\nu}{}_{,\kappa}\U^{\gamma\rho}{}_{,\sigma}  ,\label{einsta6} \\
B^{\alpha\beta} & := 4\lambda \gb_{\kappa\sigma}\left(2\gb^{\gamma ( \alpha}
\U^{\beta )\sigma}{}_{,\rho}\U^{\kappa\rho}{}_{,\gamma}-\Half\gb^{\alpha\beta}
\U^{\kappa}{\rho}{}_{\gamma}\U^{\sigma\gamma}{}_{\rho}
-\gb^{\gamma\rho}\U^{\alpha\kappa}{}_{,\gamma}\U^{\beta\sigma}{}_{\rho}\right)
 , \label{einsta7} \\
C^{\alpha \beta} & := 4\lambda^{2}\left( \U^{\alpha\beta}{}_{,\kappa}
\U^{\kappa\rho}{}_{,\rho} - \U^{\alpha\kappa}{}_{,\rho} \U^{\beta\rho}{}_{,\kappa}
\right)  ,  \label{einsta8} \\
D^{\alpha \beta} & := \gb^{\mu\nu}\U^{\alpha \beta}{}_{,\mu\nu}+
\gb^{\alpha\beta}\U^{\mu\nu}{}_{,\mu\nu} -2\U^{\mu ( \alpha}{}_{,\mu\nu}
\gb^{\beta ) \nu}  ,  \label{einsta9} 
}
and $T^{\alpha \beta}$ is the stress-energy tensor. 
Following \cite{Heil95},
we  choose harmonic coordinates
\eqn{harmonic}{
\nabla_{\alpha}\nabla^{\alpha} x^{\beta} = 0 \, , \quad \text{ or equivalently}
\quad \U^{\alpha \beta}{}_{,\beta} = 0 \, ,
}
which allows us to write the full Einstein field equations as
\lgath{reduced}{
\U^{\alpha \beta}{}_{,\beta} = 0 \, ,\label{reduced2} \\
4\pi G |\df| T^{\alpha \beta} = E^{\alpha \beta} \, ,\label{reduced3}
}
where
\lgath{reduceda}{
E^{\alpha \beta}  := \gob^{\mu\nu}\U^{\alpha\beta}{}_{,\mu\nu} + 
4\lambda^{2}\left(\U^{\mu\nu}\U^{\alpha\beta}{}_{,\mu\nu} + 
\U^{\alpha\beta}\U^{\mu\nu}{}_{,\mu\nu} - 2\U^{\mu (\alpha}{}_{,\mu\nu}
\U^{\beta ) \nu}\right) \notag \\
 + A^{\alpha\beta} + B^{\alpha\beta} + C^{\alpha\beta} \label{reduceda1} \, .
}
The equations \eqref{reduced3} will be 
called the \emph{reduced field equations}.

It is important to recognize that for $\lambda > 0$ the 
reduced field equations \eqref{reduced3}
are not equivalent to the Einstein field equations \eqref{einstA}
or equivalently \eqref{einst}.
However, it is shown in 
\cite{Heil95} \S 6  that if $T^{\alpha \beta}{}_{;\beta}=0$
 and \eqref{reduced3}
can be solved and the stress-energy tensor $T^{\alpha \beta}$ satisfies
certain conditions then the harmonic condition \eqref{reduced2} will be
automatically satisfied. In this case, a solution to \eqref{reduced3} will
actually be a solution to the full Einstein equation \eqref{einstA}.

We will let
$A = A_{\alpha} dx^{\alpha}$
denote the $SU(2)$-gauge potential and
$\psi$ the dilaton field. 
The SU(2) Yang-Mills-dilation equations are
\lgath{YMd}{
D^{\alpha}\left( e^{2\kappa \psi}F_{\alpha \beta} \right) = 0 \, ,\label{YMd1}\\
\nabla^{\alpha}\nabla_{\alpha}\psi - \frac{\kappa \ell_{Y}}{\ell_{d}}
e^{2\kappa\psi}g^{\alpha\beta}g^{\mu\nu} \ip{F_{\alpha\mu}}{F_{\beta\nu}} = 0\, , \label{YMd2}
}
where $D_{\alpha}(\cdot) := \nabla_{\alpha}(\cdot)+ [A_{\alpha},\cdot]$
is the gauge covariant derivative, $\ell_{Y}$ is the Yang-Mills coupling
constant, $\{\ell_{d}, \kappa\}$ is the dilaton coupling constants, 
\leqn{gfield}{
F_{\alpha \beta} := A_{\beta , \alpha} - A_{\alpha , \beta} +
[A_{\alpha}, A_{\beta}]
}
is the gauge field strength, and 
$\ip{\cdot}{\cdot}$ is an $\Ad$-invariant, positive definite
inner-product on $\sU{2}$. 
Multiplying \eqref{YMd1} and \eqref{YMd2} by $\sqrt{\lambda|g|}$ and
$\lambda|g|$, respectively, we find that
\lgath{Ymda}{
\gb^{\alpha\nu}\left(F_{\alpha\beta,\nu} - \Gamma^{\mu}_{\alpha\nu}
F_{\mu\beta} - \Gamma^{\mu}_{\beta\nu} F_{\alpha\mu} +
2\kappa \psi_{,\nu} F_{\alpha\beta} + [A_{\nu},F_{\alpha\beta}]\right) = 0 
\, , \label{Ymda1} \\
\gb^{\alpha\beta}\left(\psi_{,\alpha\beta} - \Gamma^{\mu}_{\alpha\beta}
\psi_{,\mu} - \frac{\kappa\ell_{Y}}{\ell_{d}}\frac{e^{2\kappa \psi}}{\sqrt{|\df|}}
\gb^{\mu\nu}\ip{F_{\alpha\mu}}{F_{\beta\nu}}\right) = 0 \, , \label{Ymda2} 
} 
where the Christoffel $\Gamma^{\alpha}_{\beta\gamma}$ symbols are given by
\leqn{Christ}{
\Gamma^{\alpha}_{\beta\gamma} = 
\gb^{\alpha\mu}(2\gb_{\beta\sigma}\gb_{\gamma\tau} - \gb_{\beta\gamma}
\gb_{\sigma\tau})\U^{\sigma\tau}{}_{,\mu} + 2\lambda (
\gb_{\sigma\tau}\delta^{\alpha}_{(\beta}\U^{\sigma\tau}{}_{,\gamma)}-
2\gb_{\sigma(\beta}\U^{\alpha\sigma}{}_{,\gamma)} ) \, .
}

The stress energy tensor is given by
\lalign{streng}{
T^{\alpha \beta} = \Half \ell_{d} &\left( g^{\alpha\mu}g^{\beta\nu}
\psi_{,\mu}\psi_{,\nu} - \Half g^{\alpha\beta}g^{\mu\nu}\psi_{,\mu}\psi_{,\nu}
\right) + \notag \\
& \ell_{Y} e^{2\kappa\psi}\left( g^{\alpha\mu}g^{\beta\nu}g^{\sigma\tau}
\ip{F_{\mu\sigma}}{F_{\nu\tau}}-\Quarter g^{\mu\nu}g^{\sigma\tau}g^{\alpha\beta}
\ip{F_{\mu\sigma}}{F_{\nu\tau}} \right) \label{streng1} \, .
}
Using the YMd equations \eqref{YMd1}-\eqref{YMd2}
, it is straightforward
to verify that any YMd solution satisfies   
\leqn{streng1a}{
T^{\alpha \beta}{}_{;\beta} = 0 
}
automatically, irrespective of the metric. Consequently,
it will be enough to solve the reduced field equations
\eqref{reduced3}
and the YMd equations \eqref{YMd1}-\eqref{YMd2}
to obtain a solution to the full EYMd field equations. 

Let
\leqn{streng2}{
\Tc^{\alpha\beta} := 4\pi G|\df| T^{\alpha\beta}
}
so that
\lalign{streng3}{
\Tc^{\alpha \beta}& = 2\pi G \ell_{d} \left( \gb^{\alpha\mu}\gb^{\beta\nu}
\psi_{,\mu}\psi_{,\nu} - \Half \gb^{\alpha\beta}\gb^{\mu\nu}\psi_{,\mu}\psi_{,\nu}
\right) + \notag \\
& 4\pi G\frac{\ell_{Y}}{\sqrt{|\df|}} e^{2\kappa\psi}\left( \gb^{\alpha\mu}\gb^{\beta\nu}
\gb^{\sigma\tau}
\ip{F_{\mu\sigma}}{F_{\nu\tau}}-\Quarter \gb^{\mu\nu}\gb^{\sigma\tau}\gb^{\alpha\beta}
\ip{F_{\mu\sigma}}{F_{\nu\tau}} \right) \label{streng4} \, .
}
 
\subsect{units}{Interpretation of solutions for varying $\lambda$}

Solving equations  
\eqref{reduced2}, \eqref{reduced3}, \eqref{YMd1}, and \eqref{YMd2}
via the Newtonian perturbation method will produce
a 1-parameter
family of solutions
\eqn{1param}{
\{g_{\alpha\beta}^{\lambda},A_{\alpha}^{\lambda},\psi^{\lambda}\}
\quad \lambda \in (0,\Lambda)}
which will solve the EYMd equations 
\lgath{EYMdAA}{
G_{\alpha\beta} = 8\pi G_\lambda T_{\alpha\beta} \quad
G_{\lambda} := \lambda^2 G \, ,\label{EYMdAA.1} \\
D^{\alpha}\left( e^{2\kappa \psi}F_{\alpha \beta} \right) = 0 \, ,
\label{EYMdAA.2}\\
\nabla^{\alpha}\nabla_{\alpha}\psi - \frac{\kappa \ell_{Y}}{\ell_{d}}
e^{2\kappa\psi}g^{\alpha\beta}g^{\mu\nu}
\ip{F_{\alpha\mu}}{F_{\beta\nu}} = 0\,\label{EYMdAA.3} .
}
The maximum interval $(0,\Lambda)$ for which the one parameter
family of solutions is defined will, in general, depend
on the equations and the ``Newtonian
solution'' (i.e. the singular solution at $\lambda=0$) 
that is used to start the perturbation argument. We will
not discuss methods in this paper to estimate the size of $\Lambda$
and therefore will have to consider the size of $\Lambda$
as unknown.

Equation \eqref{EYMdAA.1} shows that the limit $\lambda \rightarrow 0$
is equivalent to the limit that the gravitational coupling
constant $G_{\lambda} \rightarrow 0$. 
However, this
is not the only interpretation of the limit $\lambda \rightarrow 0$.
Rescaling the fields as follows
\eqn{solBB}{
g_{\alpha\beta} := \lambda^{-2} g_{\alpha\beta}^{\lambda}\, ,
\quad A_{\alpha} := A^{\lambda}_\alpha\, ,\quad \text{and} \quad 
\psi := \lambda\psi^{\lambda} } 
shows that $\{g_{\alpha\beta},A_\alpha,\psi\}$ solve
the EYMd equations \eqref{EYMdAA.1}-\eqref{EYMdAA.3}
with the following change of coupling constants
\eqn{coupling}{
G_{\lambda} \mapsto G\, , \quad \ell_Y \mapsto \ell_Y,
\quad \ell_d \mapsto \ell_d, \quad \text{and}\quad
\kappa \mapsto \kappa/\lambda\, .
}
Thus the limit $\lambda \rightarrow 0$
can be also interpreted as the limit that the dilaton
coupling constant $\kappa \rightarrow \infty$. We
conclude there does not exist a unique interpretation of the
limit $\lambda \rightarrow 0$ and moreover only
certain variables will be defined in the limit
$\lambda \rightarrow 0$. In our case, the variables that 
continue to be defined at $\lambda = 0$ are
the unscaled dilaton field $\psi$, the gauge potential $A_{\alpha}$,
and the metric density $\U_{\alpha\beta}$. We stress
that the metric $g_{\alpha\beta}$ does not exist
at $\lambda = 0$.

Since we do not know the size of $\Lambda$, the Newtonian
perturbation method does
not necessarily produce solutions for all possible values
of the coupling constants. Consider $\{G,\ell_d,\kappa,\ell_Y\}$
as a set of coupling constants in some fixed units for which we
would like to have a solution to the EYMd equations. If $\Lambda > 1$,
then we could choose $\lambda =1$ and in that case $G_{\lambda} = G$
and we would have a solution to the EYMd equations for the
fixed coupling constants $\{G,\ell_d,\kappa,\ell_Y\}$. On the other
hand, if $\Lambda < 1$ then $G_{\lambda} < G$ and we will have
to be satisfied with a solution to the EYMd equation where the
gravitational coupling constant is smaller than $G$. As discussed
above, we could rescale the metric and the dilaton field to
get a solution where $G_{\lambda} = G$ provided that we
change the dilaton coupling constant $\kappa$ to $\kappa/\lambda$.
Thus in general the solutions that the Newtonian perturbation
methods produces will have some restriction on the size of
at least one of the coupling constants.

%% file: wsob.tex
\sect{wsob}{Weighted Sobolev Spaces}

Let $V$ denote a finite dimensional vector space with
norm $\enorm{\cdot}$. 
\begin{Def}  \label{wsobdef1} \mnote{[wsobdef1]}
The \emph{weighted Lesbegue space} $\text{L}^{p}_{\delta}(\Rbb^{n},V)$,
$1\leq p \leq \infty$, with weight $\delta \in \Rbb$ is the set
of all measurable maps from $\Rbb^{n}$ to $V$ in 
$\text{L}^{p}_{\text{loc}}(\Rbb^{n},V)$ such that the norm
\eqn{wLebnorm}{
\norm{u}_{p,\delta} = \left\{ \begin{array}{ll}
{\displaystyle \left(\int_{\Rbb^{n}}|u|^{p}\sigma^{-\delta p -n} d^{n}x\right)^
{\frac{1}{p}}} &
\text{if $p <\infty$}\, \\
&  \\
\text{\emph{ess sup}}_{\Rbb^{n}}(\sigma^{-\delta}|u|) & \text{if $p=\infty$} \, ,
\end{array} \right.
}
is finite. Here $\sigma(\xb) := \sqrt{|x|^{2} + 1}$. If $V=\Rbb$
then we write $\text{L}^{p}_{\delta}(\Rbb^{n})$ instead
of $\text{L}^{p}_{\delta}(\Rbb^{n},V)$.
\end{Def}

\begin{Def}  \label{wsobdef2} \mnote{[wsobdef2]}
The \emph{weighted Sobolev space} $\W^{k,p}_{\delta}(\Rbb^{n},V)$,
$1\leq p \leq \infty$, $k\in\Nbb_{0}$, with weight $\delta \in \Rbb$ is the set
\eqn{Sobelev}{
\W^{k,p}_{\delta}(\Rbb^{n},V):= \{\, u \in 
\text{L}^{p}_{\delta}(\Rbb^{n},V) \,|\,
\partial^{I}u\in \text{L}^{p}_{\delta-|I|}(\Rbb^{n},V) \;\text{for all
$I : |I|\leq k$}\,\}
}
with norm
\eqn{wSobnorm}{
\norm{u}_{k,p,\delta} := \sum_{|I|\leq k} \norm{\partial^{I}u}_{p,\delta-|I|}\, ,
}
where $I=(I_{1},I_{2},\ldots,I_{n})$ is a multi-index
and $\partial^{I} :=\partial_{1}^{I_{1}}\partial_{2}^{I_{2}}\cdots\partial_{n}^{I_{n}}$.
If $V=\Rbb$ then we will write $\W^{k,p}_{\delta}(\Rbb^{n})$ instead
of $\W^{k,p}_{\delta}(\Rbb^{n},V)$.
\end{Def}
From the definition, it is clear that differentiation
\leqn{diff}{
\partial_{j} : \W^{k,p}_{\delta}(\Rbb^{n},V) \rightarrow
\W^{k-1,p}_{\delta-1}(\Rbb^{n},V)
}
is a continuous linear map. Also from the definition and H\"{o}lders
inequality it is easy to show (see also \cite{Bart86}, proposition
1.2 (i) )  that if $k_{1} \geq k_{2}$ and
$\delta_{1} \leq \delta_{2}$ then
\leqn{embedA}{
\W^{k_{1},p}_{\delta_{1}}(\Rbb^{n},V) \subset 
\W^{k_{2},p}_{\delta_{2}}(\Rbb^{n},V) \, .
}
Finally, we note that the set $\Co(\Rbb^{n},V)$ of smooth maps
from $\Rbb^{n}$ to $V$ with compact support is dense in $\W^{k,p}_{\delta}(\Rbb^{n},V)$. 
As above, if $V=\Rbb$ then we write $\Co(\Rbb^{n})$ instead of $\Co(\Rbb^{n},V)$.
We will now state some results in weighted Sobolev spaces that
will be needed. For proofs
see \cite{Bart86} and \cite{CHCH81}.  
\begin{lem}  \label{multiply} \mnote{[multiply]}
If there exists a multiplication 
$V_{1}\times V_{2} \rightarrow V_{3}$ $(u,v)\mapsto u\cdot v$ then 
the corresponding multiplication
\eqn{multiply1}{
\W^{k_{1},p}_{\delta_{1}}(\Rbb^{n},V_{1})\times \W^{k_{2},p}_{\delta_{2}}(\Rbb^{n},V_{2}) 
\rightarrow \W^{k_{3},p}_{\delta_{3}}(\Rbb^{n},V_{3})\; : \; (u,v) 
\mapsto u\cdot v
}
is bilinear and continuous if
$k_{1},k_{2} \geq k_{3}$, $k_{3} < k_{1} + k_{2} - n/p$, and
$\delta_{1} +\delta_{2} < \delta_{3}$ .
\end{lem}
\begin{proof} See lemma 2.5 in \cite{CHCH81} for the case $p=2$. For all
$p$ this can be proved easily using theorem 1.2 of \cite{Bart86}.
\end{proof}

\begin{thm}  \label{laplace} \mnote{[laplace]}
For $\delta < 0$ the Laplacian
\eqn{laplace1}{
\Delta := \sum_{j=1}^{n}\partial^2_j : 
\W^{k,p}_{\delta}(\Rbb^{n},V) \rightarrow
\W^{k-2,p}_{\delta-2}(\Rbb^{n},V)
}
is continuous and injective. Moreover if
$2-n <\delta < 0$ then the Laplacian is
an isomorphism. The inverse is given by
\leqn{laplace2}{
(\Delta^{-1}u)(\xb) = \frac{1}{(2-n)\omega_{n} }\int_{\Rbb^{n}} \frac{u(\yb)}{|\xb-\yb|^{(n-2)}}
d^{n}y\, , 
}
where $\omega_{n}$ is the area of the unit sphere in $\Rbb^{n}$.
\end{thm}

\begin{lem} \label{embed} \mnote{[embed]}
For $k_{1} > k_{2}$, $\delta_{1} < \delta_{2}$, and $1\leq p < \infty$
the embedding\\ $\W^{k_{1},p}_{\delta_{1}}(\Rbb^{n},V) 
\rightarrow \W^{k_{2},p}_{\delta_{2}}(\Rbb^{n},V)$ is 
compact.
\end{lem}

As with the Sobolev spaces, we can define a weighted versions of 
the $C^{k,\alpha}(\Rbb^{n},V)$ spaces.
For a map $u\in C^{0}(\Rbb^{3},V)$ and $\delta\in \Rbb$, $\alpha > 0$ , let
\eqn{rsobdef3.2}{
\norm{u}_{C^{0,\alpha}_{\delta}}  := \norm{u}_{C^{0}_{\delta}}
+ \sup_{x\in \Rbb^{n}}\Bigl(\sigma^{-\delta+\alpha}(x) \sup_{4|x-y|\leq \sigma(x
)}\frac{|u(x)-u(y)|}{|x-y|^{\alpha}}
\Bigr)\, .
}
Using this norm  we define the norm
$\norm{\cdot}_{C^{k,\alpha}_{\delta}}$ in the usual way:
\eqn{rsobdef3.4}{
\norm{u}_{C^{k,\alpha}_{\delta}} := \sum_{|I|\leq k} \norm{\partial^{I}u}_{C^{0,
\alpha}_{\delta-|I|}} \, .
}
So then
\eqn{rsobdef3.6}{
 C^{k,\alpha}_{\delta}(\Rbb^{n},V) :=
\bigl\{\, u \in C^{k}(\Rbb,V) \, | \, \norm{u}_{C^{k,\alpha}_{\delta}} < \infty
\: \bigr\}\, .
}

%% file: sss.tex
\sect{sss}{Static spherically symmetric fields}

We assume that all the  fields are static and that $\partial_{0}$
is a timelike hypersurface orthogonal Killing vector field for the
metric. Therefore
$\partial_{0} \U^{\alpha \beta} = 0$, $\partial_{0} A_{\alpha} = 0$,
$\partial_{0} \psi = 0$,
and $\U^{j 0} = \U^{0 j} = 0$.
Since $\U^{\alpha \beta}$ is symmetric, i.e. $\U^{\alpha \beta} = 
\U^{\beta \alpha}$, we define the following subspace of
the 4 by 4 matrices
\eqn{static3}{
\Sbb := \{\, X=(X^{\alpha \beta}) \in \Mbb_{4\times 4} \,|\, X^{\alpha \beta}
= X^{\beta \alpha} \;\text{and}\; X^{0 j} = 0 \,\}\;.
}
Then letting $\U = (\U^{\alpha\beta})$, $\U$ takes values
in $\Sbb$. We will also assume that $A_{0}=0$.
Therefore 
if we write the gauge potential $A_{i}$ as a 3-tuple
$A = (A_{1},A_{2},A_{3})$ then the gauge potential
$A$ takes values in the space $\sU{2}^{3}$ which
carries a norm $\enorm{A}^{2} := \sum_{i=1}^{3} \ip{A_{i}}{A_{i}}$.
Therefore  $\W^{k,p}_{\delta}(\Rbb^{3},\Sbb)$, 
$\W^{k,p}_{\delta}(\Rbb^{3})$ and $\W^{k,p}_{\delta}(\Rbb^{3},\sU{2}^{3})$
are appropriate functions spaces for the static 
metric densities, dilaton fields,
and gauge potentials, respectively.   

In addition to being static, we will also assume that our fields 
are spherically symmetric.
To define what we mean by spherical symmetry we first
need to specify an action of $SO(3)$ on spacetime $\Rbb^{4}$.
We want $SO(3)$ to act on the hypersurfaces
orthogonal to the timelike killing vector field $\partial_{0}$.
So using the matrix representation of $SO(3)$ given by
$SO(3) = \{\, a\in \Mbb_{3\times 3}\,|\, \text{$a^{t} = a^{-1}$ and
$\det(a)=1$} \,\}$
we define a $SO(3)$ action on spacetime by
$\Phi : SO(3) \times \Rbb^{4} 
\rightarrow \Rbb^{4} \; :\; (a,(x^{0},\xb)) \rightarrow
\Phi_{a}(x^{0},\xb):=(x^{0},a\xb)$
where we are treating $\xb$ as a column vector and $a\xb$ denotes matrix
multiplication. We then get the induced action on functions via pullbacks.
Therefore $SO(3)$ acts on the dilaton field $\psi(\xb)$
as follows
$\Phi_{a}(\psi)(\xb) :=  \psi(a^{\text{t}}\xb)$.
Lifting the $SO(3)$ action on spacetime to the tensor bundle, we
get the following action on the metric densities
$\Phi_{a}(\U)(\xb) := \at \U(a^{\text{t}}\xb)\at^{\text{t}}$
where
$\at := \text{diag}(1,a)$.
Let $\sCo(\Rbb^{3})$ denote the set of smooth $SO(3)$-invariant functions with
compact support, i.e.
$\sCo(\Rbb^{3}) := \{\, \psi \in \Co(\Rbb^{3})\, | \, \psi = 
\Phi_{a}\psi \;\text{for all} \; a\in SO(3)\,\}$.
In other words, $\sCo(\Rbb^{3})$ is the set of radial functions
on $\Rbb^{3}$. 
Similarly, define
\eqn{cssdens}{
\sCo(\Rbb^{3},\Sbb) := \{\, \U \in \Co(\Rbb^{3},\Sbb) \, | \,
\U = \Phi_{a} \U \; \text{for all}\; a \in SO(3) \, \}\,.
}

In addition to being spherically symmetric, we will
assume that our gauge potential is purely magnetic.
Choosing an appropriate gauge, the gauge potential
can then be written as \cite{k5109}
\eqn{sgpot}{
A_{i}(\xb) := u(\xb)\epsilon_{i}{}^{j}{}_{k} x^{k}\tau_{j}
}
where $u(\xb) = u(|\xb|)$ and 
\eqn{su2basis}{
\tau_{1} = \frac{1}{2i}\begin{pmatrix} 0 & 1\\ 1& 0 \end{pmatrix}\, , \;
\tau_{2} = \frac{1}{2i}\begin{pmatrix} 0 & -i\\ i& 0 \end{pmatrix}\, , \;
\tau_{3} = \frac{1}{2i}\begin{pmatrix} 1 & 0\\ 0& -1 \end{pmatrix}\, ,
}
is a basis for $\sU{2}$. This form of the gauge potential
is known as the Witten ansatz.

We then define the  set of smooth static spherically symmetric 
purely magnetic gauge potentials with compact support
by
\eqn{cssgpot}{
\sAo := \{ A : \Rbb^{3} \rightarrow 
\sU{2}^{3}\, | \, A_{i}(\xb) = 
u(\xb)\epsilon_{i}{}^{j}{}_{k} x^{k}\tau_{j}\; \text{for some} \;u\in \sCo(\Rbb^{3})
\, \} \, .
}
Notice that every $A \in \sAo$ satisfies
\leqn{divA1}{
\Div A := \sum_{j=1}^{3} \partial_{j} A_{j} = 0 \; .
}
So then the spherically symmetric Sobolev spaces we consider
are
\lgath{sspace}{
\Dc^{k,p}_{\delta}:= \overline{\sCo(\Rbb^{3})} \subset
\W^{k,p}_{\delta}(\Rbb^{3}) \, ,  \label{sspace1} \\
\Uc^{k,p}_{\delta}:= \overline{\sCo(\Rbb^{3},\Sbb)} \subset
\W^{k,p}_{\delta}(\Rbb^{3},\Sbb) \, , \label{sspace2} \\
\intertext{and}
\Ac{}^{k,p}_{\delta} := \overline{\sAo} \subset 
\W^{k,p}_{\delta}(\Rbb^{3},\sU{2}^{3}) \, . \label{sspace3}
}
Because of \eqref{divA1} we have
\leqn{divA2}{
\Div A = 0 \quad \text{for all $A \in \Ac{}^{k,p}_{\delta}$}
}
by the density of $\sAo$ in $\Ac{}^{k,p}_{\delta}$ and
the continuity of differentiation (see \eqref{diff}). Therefore
each $A\in \Ac{}^{k,p}_{\delta}$ automatically satisfies the
Coulomb gauge condition.

We now analyze the Laplacian $\Delta = \sum_{j=1}^{3}\partial_j^2$
on the spherically symmetric spaces \eqref{sspace1}-\eqref{sspace3}.
\begin{prop} \label{slaplaceA} \mnote{[slaplaceA]}
For $-1 <\delta < 0$ the Laplacian
$\Delta : \Dc^{k,p}_{\delta} \rightarrow
\Dc^{k,p}_{\delta-2}$ is an isomorphism.
\end{prop}
\begin{proof}
Straightforward calculation shows that 
$\Delta(\sCo(\Rbb^{3})) \subset \sCo(\Rbb^{3})$.
Using formula \eqref{laplace2}, it is not difficult to verify that
if $\psi \in \sCo(\Rbb^{3})$ then $\Phi_{a}(\Delta^{-1}\psi) = \Delta^{-1}\psi$
for all $a\in SO(3)$. The proposition then follows from these two results
and theorem \ref{laplace}.
\end{proof}
The next proposition is proved in the same fashion.
\begin{prop} \label{slaplaceB} \mnote{[slaplaceB]}
For $-1 <\delta < 0$ the Laplacian
$\Delta : \Uc^{k,p}_{\delta} \rightarrow
\Uc^{k-2,p}_{\delta-2}$
is an isomorphism.
\end{prop}
We will often use the following notation
\eqn{rdef}{
r := |\xb| \quad \text{and} \quad (\cdot)' = \frac{d(\cdot)}{dr} \, .
}
The next proposition is interesting because it shows
that on the space of the spherically symmetric gauge
potentials, the Laplacian is invertible for a larger
range of weights $\delta$ than one would expect
from theorem \ref{laplace}. The reason for this is that the
Laplacian (see equation \eqref{slaplaceC3} below) acting
on the space of spherically symmetric gauge potentials
is essentially equivalent to the Laplacian acting on
the space of spherically symmetric functions on $\Rbb^5$.
We note that this observation has also been used in
\cite{GlaStr} to construct global solutions of the Yang-Mills
equations on Minkowski spacetime.
\begin{prop} \label{slaplaceC} \mnote{[slaplaceC]}
For $-2 <\delta < 1$, $\delta \neq -1, 0$, the Laplacian
$\Delta : \Ac^{2,p}_{\delta} \rightarrow
\Ac^{0,p}_{\delta-2}$
is an isomorphism.
\end{prop}
\begin{proof}  
By definition of $\sAo$, if $A\in \sAo$ then 
$A_{i} = u(r)\epsilon_{i}{}^{j}{}_{k} x^{k}\tau_{j}$
for some $u\in \sCo(\Rbb^{3})$. So 
\leqn{slaplaceC3}{
\Delta A_{i} = \left( u''(r) + \frac{4}{r}u'(r)\right)\epsilon_{i}{}^{j}{}_{k}
x^{k} \tau_{j}
}
and hence $\Delta(\sAo)\subset \sAo$. Therefore  
$\Delta : \Ac^{2,p}_{\delta}$ $\rightarrow$ $\Ac^{0,p}_{\delta-2}$
is continuous by the density of $\sAo$ and the continuity of
$\Delta : \W^{2,p}_{\delta}(\Rbb^{3},\sU{2}^{3}) \rightarrow
\W^{0,p}_{\delta-2}(\Rbb^{3},\sU{2}^{3})$. 

Suppose $A \in \Ac^{2,p}_{\delta}$ satisfies $\Delta A = 0$. Then
by elliptic regularity $A \in C^{\infty}$ and hence
$A_{i} = u(r)\epsilon_{i}{}^{j}{}_{k} x^{k}\tau_{j}$ for
some smooth function $u(r)$ on $[0,\infty)$ that satisfies
(i) $u(r) = u_{0} + \Ord(r^2)$ as $r\rightarrow 0$ for some constant $u_{0}$
and (ii) the differential equation
\eqn{slaplaceC4}{
\bar{u}''(r) + \frac{4}{r}\bar{u}'(r) = 0 \, .
}
However, the general solution to this equation is
$\bar{u}(r) = c_{1} + c_{2}r^{-3}$
for some constants $c_{1}$, $c_{2}$. This shows
that $u(r) = u_{0}$ as $u(r)$ is bounded near $r=0$. So
$A_{i}(x) = u_{0} \epsilon_{i}{}^{j}{}_{k} x^{k}\tau_{j}$.
Any positive definite invariant product $\ip{\cdot}{\cdot}$ on
$\sU{2}$ is given by $\ip{A}{B} = -2\alpha \text{Tr}(AB)$
for some $\alpha > 0$. A short calculation then shows that 
$|\epsilon_{i}{}^{j}{}_{k}x^{k}\tau_{j}|^2= 2\alpha r^2$.
Using this we find that $|A(x)| = \sqrt{2\alpha}|u_{0}|r$. This shows
that $A\in \Ac^{2,p}_{\delta}$ for $\delta < 1$ only
if $u_{0} = 0$. This establishes that $\ker\Delta|_{\Ac^{2,p}_{\delta}}
=0$ for $\delta < 1$.

It follows from theorem 1.10 in \cite{Bart86} that $\Delta : 
\W^{2,p}_{\delta}(\Rbb^{3},\sU{2}^{3}) \rightarrow
\W^{0,p}_{\delta-2}(\Rbb^{3},\sU{2}^{3})$ has closed range for
$-2 < \delta < 1$ and $\delta \neq -1,0$.
This implies that $\Delta : \Ac^{2,p}_{\delta} \rightarrow 
\Ac^{0,p}_{\delta}$ also has closed range for the same values
of $\delta$. With respect to the pairing $(A,B) = \int \ip{A}{B}d^{3}x$
the Laplacian has a formal adjoint $\Delta^{*}=\Delta$. Since
$W^{0,p}_{\delta-2}(\Rbb^{3},\sU{2}^{3})^{*}=
W^{0,p'}_{-1-\delta}(\Rbb^{3},\sU{2}^{3})$
where $p'=p/(p-1)$, it follows from
propositions 1.6 and   1.14 of \cite{Bart86} that
$\text{ker}{\Delta^{*}}\subset W^{2,p}_{-1-\delta}(\Rbb^{3},\sU{2})$.
Therefore the arguments in the previous paragraph show that 
\leqn{slaplaceA.11a}{
\dim\text{coker} \Delta\bigl|_{\Ac^{2,p}_{\delta}} =
\dim \text{ker} \Delta\bigl|_{\Ac^{2,p}_{-1-\delta}} = 0
}
for $-2< \delta <1$. Hence $\Delta$ is an isomorphism
for $\delta \neq 0,-1$ and $-2 < \delta <1 $.
\end{proof}

%% file: newt.tex
\sect{newt}{Yang-Mills-dilaton solutions}

To employ the Newtonian perturbation method, we need
static solutions to 
the Euclidean YMd equations
\lgath{YMdI}{
\Delta \alpha -
\frac{\kappa\ell_{Y}}{\ell_{d}}e^{2\kappa\alpha}\delta^{ij}\delta^{kl}\ip{F^{W}_
{ik}}
{F^{W}_{jl}} = 0 \label{YMdI1}\, , \\
\delta^{ik}\big(\partial_{k}F^{W}_{ij}
+ 2\kappa F^{W}_{ij}\partial_{k}\alpha + 
[W_{k},F^{W}_{jk}]\big) = 0\label{YMdI2}\, . 
}
Assuming that $\alpha$ is a function of $r$ only and
\leqn{Wsplit}{
W_{i}(r) := \frac{w(r)-1}{r^2}\epsilon_{i}{}^{j}{}_{k}x^{k}\tau_{j} \, ,
}
the YMd equations \eqref{YMdI1}-\eqref{YMdI2} become
\lgath{YMd.1}{
w'' = -2\kappa \alpha'w' + \frac{(w^{2}-1)w}{r^{2}} \, , \label{YMd.1.1} \\
(r^{2}\alpha')' = \frac{4\kappa\ell_{Y}}{\ell_{d}}e^{2\kappa\alpha}
\left( {w'}^{2} + \frac{(w^{2}-1)^{2}}{2 r^{2}}
\right)
\, .
\label{YMd.1.2}
}
It is easy to check that
\eqn{YMd.2}{
 \bar{w}(r) := w\left(
\frac{\sqrt{8}\kappa\sqrt{\ell_Y}}{\sqrt{\ell_d}} r\right)
\quad \text{and} \quad \bar{\alpha}(r) := 2\kappa\alpha\left(
\frac{\sqrt{8}\kappa\sqrt{\ell_Y}}{\sqrt{\ell_d}} r\right)
}
satisfy \eqref{YMd.1.1}-\eqref{YMd.1.2} with
$\kappa =1/2$ and $4\kappa \ell_{Y}/\ell_{d} = 1$.
Therefore, we can, without any loss of generality, consider the
equations
\lgath{YMd.3}{
w'' = -\alpha'w' + \frac{(w^{2}-1)w}{r^{2}} \, , \label{YMd.3.1} \\
(r^{2}\alpha')' = e^{\alpha}
\left( {w'}^{2} + \frac{(w^{2}-1)^{2}}{2 r^{2}}
\right)
\, .
\label{YMd.3.2}
}
We note that these equations have a scaling symmetry. To be 
precise, if $(w(r),\alpha(r))$ is a solution
then
\leqn{isoB6}{
w_{\beta}(r) := w(e^{\beta/2}r) \quad \alpha_{\beta}:=\alpha(e^{\beta/2}r)-\beta
}
will also solve the equations for any $\beta \in \Rbb$.

The next theorem provides the existence of an infinite number of
solutions to \eqref{YMd.3.1}-\eqref{YMd.3.2}. 
\begin{thm}{\emph{[theorem 1,\cite{HMT95}]}} \label{existA} \mnote{[existA]}
There exists a sequence $n=1,2,3,\ldots$ of analytic 
solutions $(w_{n}(r),\alpha_{n}(r))$ to the YMd equations
\eqref{YMd.3.1}-\eqref{YMd.3.2} defined on $(0,\infty)$ such that
$w_{n}$ has precisely $n$ zeros and
$\lim_{r\rightarrow\infty} w_{n}(r) = (-1)^{n}$.
\end{thm}
\begin{rem} \label{existAA} \mnote{[existAA]}
It is also established in \cite{HMT95} that the solutions $(w_{n}(r),\alpha_{n}(r))$
from theorem \ref{existA} satisfy the following
\begin{enumerate}
\item $\lim_{r\rightarrow\infty}\alpha_{n}(r) = \text{const}$,
\item  $|w|\leq 1$, $w_{n}' \in \emph{\text{o}}(r^{-1})$ and
$\alpha_{n}' = \emph{\text{O}}(r^{-2})$ as $r\rightarrow \infty$,
\item $w_{n}(r)$ and $\alpha_{n}(r)$ are analytic in a neighborhood of $r=0$ and
\eqn{existA1}{
w_{n}=1-\beta_{n} r^2 + \emph{\text{O}}(r^{4}) 
\quad\text{ as $r\rightarrow 0$}
}
for a constant $\beta_{n} > 0$.
\item $w_{n}'$ is either strictly positive or negative for $r$ large enough.
\end{enumerate}
By using the scaling transformation \eqref{isoB6}, we can assume
$\lim_{r\rightarrow\infty} \alpha_{n}(r) = 0$ .
\end{rem}

These are the solutions that we will use to start our perturbation
argument. However, for these solutions to be useful for our purposes
we will need more information about their large $r$ behavior.
The required information is contained in the next proposition.

\begin{prop} \label{existC} \mnote{[existC]}
Suppose $(w(r),\alpha(r))$ is a solution to the flat YMd equations 
\eqref{YMd.3.1}-\eqref{YMd.3.2} defined on $(0,\infty)$ that 
satisfies $|w(r)| < 1 $ for all $r\in (0,\infty)$, $\lim_{r \searrow 0}w'(r)=0$,
$\lim_{r\rightarrow \infty}w(r) = 1$ or $\lim_{r\rightarrow \infty}w(r) = -1$, $\lim_{r\rightarrow\infty} \alpha(r) = 0$,
$w' = \emph{\text{o}}(1/r)$ and  
$\alpha' = \emph{\text{O}}(r^{-2})$ as $r\rightarrow \infty$. Furthermore,
suppose that there exist a $R > 0$ such that $w'(r),w(r) > 0$ or $w'(r),w(r) < 0$
for all $r\geq R$. Then for any $\epsilon \in (0,1)$:
$w''  = \emph{\text{O}}(r^{-2\epsilon-1})$, $w'  = 
\emph{\text{O}}(r^{-2\epsilon})$,
$\text{$w-1$ or $w+1$}   = \emph{\text{O}}(r^{-2\epsilon+1})$, $\alpha''  = \emph{\text{O}}(r^{-3})$,
and  $\alpha  = \emph{\text{O}}(r^{-1})$
as $r\rightarrow \infty$.
\end{prop}
\begin{proof}
Define
\eqn{existC1}{
u := 1-w^{2} \; 
}
and
\eqn{existC2}{
Z_{\pm} := \frac{1-w^{2}}{r} - 2ww'\pm\frac{w'}{r^{1/2}}\, .
}
Note that
\eqn{existC2b}{
0 < u(r)  \leq 1\; \quad \forall \; r\in (0,\infty)\, 
}
as $|w| < 1$ on $(0,\infty)$.
Also note that $Z$ can be written as
\eqn{existC3}{
Z_{\pm} = \frac{u}{r} +  u' \pm \frac{w'}{r^{1/2}}\, .
}

\begin{lem} \label{existD} \mnote{[existD]}
If $w'(r),w(r) > 0$ $(w'(r),w(r) < 0)$ for $r \geq R$ and 
there exist a $R^{*} \geq R$ such that $Z_{+}(r)<0$ 
$(Z_{-}(r)<0)$ for all $r\geq R^{*}$ then
$w'= \emph{\text{O}}(r^{-2})$ as $r\rightarrow \infty$
.
\end{lem}
\begin{proof}
We only prove the case where $w'(r)$ and $w(r)$ are both positive for sufficiently large $r$. 
The other case can be handled using
similar arguments.
Since $w(r) > 0$ for $r\geq R$,  $Z_{+}(r) < 0$ for  $r\geq R^{*}$ implies that
\eqn{existC4}{
\frac{1}{r} \leq -\frac{u'}{u} \quad \forall \; r\geq R^{*} \, 
}
as $u > 0$. Integrating this expression between $R^{*}$ and $r$ yields
\eqn{existC5}{
\ln\left(\frac{r}{R^{*}}\right) < \ln\left(\frac{u(R^{*})}{u(r)}\right) \, ,
}
or equivalently
\leqn{existC6}{
u(r) < \frac{C}{r} \quad \forall \; r\geq R^{*} \, 
}
where $C = u(R^{*})\sqrt{R^{*}}$. Note that \eqref{YMd.3.1} can be written
$(e^{\alpha}w')' = -r^{-2}wu$.
Then for $r\geq R^{*}$, integration yields
\alin{existC8}{
e^{\alpha(r)}w'(r)  &= \int_{r}^{\infty} \frac{wu}{\rho^{2}} d\rho && \text{(since
$\lim_{r\rightarrow \infty} e^{\alpha(r)}w'(r) = 0$)}\\
&\leq \int_{r}^{\infty} \frac{C}{\rho^{3}}d\rho && \text{(by \eqref{existC6} and 
$|w|\leq 1$)}\\
& = \frac{2C}{3} \frac{1}{r^{2}}\, .
}
The result then follows since $w'(r) > 0$ for $r\geq R$ and $\lim_{r\rightarrow\infty}
\alpha(r) = 0$.
\end{proof}

\begin{lem} \label{existE} \mnote{[existE]}
If $w'(r),w(r) > 0$ $(w'(r),w(r) < 0)$ for $r \geq R$ and
there exists a $R^{*}\geq R$ such that $Z_{+}(r) > 0$ $(Z_{-}(r) > 0)$for all $r\geq R^{*}$ then
for any $\epsilon \in (0,1)$ $w'= \emph{\text{O}}(r^{-2\epsilon})$. 
\end{lem}
\begin{proof}
Again, we only prove the case where $w'(r)$ and $w(r)$ are both positive for sufficiently large $r$,
with the other cases following from similar arguments.
Since $\lim_{r\rightarrow \infty}w(r)=1$ there exists a $\tilde{R}\geq R^{*}$ such
that $w(r)>0$ for all $r\geq \tilde{R}$. Therefore, $Z_{+}(r) > 0$ for $r\geq R^{*}$
implies that
\eqn{existC9}{
\frac{w(1-w^{2})}{r^{2}} - \frac{2w^{2}w'}{r} + \frac{ww'}{r^{3/2}} > 0 \quad
\forall \; r\geq \tilde{R}
}
as $w'>0$ for all $r\geq R$. It then follows from \eqref{YMd.3.1} that
\eqn{existC10}{
-w'' - \alpha'w' > \frac{2w^{2}w'}{r} - \frac{ww'}{2r^{3/2}} \quad 
\forall \; r\geq \tilde{R} \; .
}
Fix $\epsilon \in (0,1)$. As $\lim_{r\rightarrow \infty}w(r) = 1$, there exists
a $R_{\epsilon} \geq \tilde{R}$ such that $w(r) \geq \sqrt{\epsilon}$ for all
$r\geq R_{\epsilon}$. Thus
\leqn{existC11}{
-w'' - \alpha'w' > \frac{2\epsilon}{r}w' - \frac{w'}{2r^{3/2}} \quad \;
\forall \; r\geq R_{\epsilon}\, .
}
Note that in deriving this inequality we have also used  $|w|\leq 1$.
Dividing \eqref{existC11} by $w'$ yields
\eqn{existC12}{
-\frac{w''}{w'} >  - \alpha' +  \frac{2\epsilon}{r} - \frac{1}{2r^{3/2}} \quad \;
\forall \; r\geq R_{\epsilon}\, .
}
Integrating gives
\eqn{existC13}{
\ln\left(\frac{w'(R_{\epsilon})}{w'(r)}\right) > \alpha(r)-\alpha(R_{\epsilon})
+ \ln\left(\left(\frac{r}{R_{\epsilon}}\right)^{2\epsilon}\right) -
\frac{1}{\sqrt{R_{\epsilon}}} \quad \forall \; r\geq R_{\epsilon}\, , 
}
and hence
\eqn{existC14}{
w'(r) < \left( \frac{w'(R_{\epsilon})e^{\alpha(r)}
e^{2 R_{\epsilon}^{-1/2}}}{e^{\alpha(R_{\epsilon})}}\right) \frac{1}{r^{2\epsilon}}
\quad \forall \; r\geq R_{\epsilon}\, .
}
The proof then follows as $\lim_{r\rightarrow \infty} \alpha(r) = 0$ and
$w'(r) > 0$ for all $r \geq R_{\epsilon}$.
\end{proof}

\begin{lem} \label{existF} \mnote{[existF]}
\eqn{existC15}{
w' = \emph{\text{O}}(r^{-2\epsilon})\quad \text{for any
$\epsilon \in (0,1)$.}
}
\end{lem}
\begin{proof}
We need to consider two cases, namely $w'(r),w(r) > 0$ and $w'(r),w(r) < 0$ for 
$r \geq R$. We will prove the lemma assuming that $w'(r),w(r) > 0$ for $r\geq R$
with the other case following from similar arguments.
We may assume that there exists
a sequence $\{r_{n}\}_{n=1}^{\infty}$ such that $R\leq r_{1} < r_{2} < r_{3} <\ldots$,
$\lim_{n\rightarrow \infty}r_{n}=\infty$, and 
$Z_{+}(r_{n}) = 0 \quad n=1,2,3,\ldots$
because otherwise we are done by  lemmas \ref{existD} and \ref{existE}. From \eqref{YMd.3.1},
it is easy to verify that $u=1-w^{2}$ satisfies
\leqn{existC17}{
u'' = \frac{2w^{2}}{r^{2}}u - 2 |w'|^{2} + 2w\alpha'w' \; .
}
Define
\leqn{existC18}{
f(r) := 2|w'|^2+\frac{wu}{r^{2}} + \left(\frac{3}{2r^{3/2}} + 
\frac{\alpha'}{r^{1/2}} - 2w\alpha'\right)w' \; .
}
Since $\alpha'= \text{O}(r^{-2})$, $|w|\leq 1$, and $w'(r) > 0$ for all $r\geq R$,
there exists a $\tilde{R} \geq R$ such that
\leqn{existC19}{
f(r) > 0 \quad \forall \; r\geq \tilde{R}.
} 
Choose $m\in \mathbb{N}$ large enough so that 
\leqn{existC19a}{
r_{m} \geq \tilde{R}\, .
}
By definition
of the $r_{n}$ we have
\leqn{existC20}{
Z_{+}(r_{m}) = \frac{u(r_{m})}{r_{m}} + u'(r_{m}) +\frac{w'(r_{m})}{r_{m}} = 0 \; .
}
Consider the following initial value problem
\lgath{existC21}{
v'' = \frac{2}{r^{2}}v+\frac{3w'}{2r^{3/2}} + \frac{\alpha'w'}{r^{1/2}}
+ \frac{wu}{r^{2}} \, , \label{existC21.1}\\
v(r_{m})=u(r_{m}) \quad v'(r_{m})=u'(r_{m})\label{existC21.2} \; .
}
From \eqref{existC17} and \eqref{existC21.1} we see that
\leqn{existC22}{
(v-u)'' = \frac{2}{r^{2}}(v-w^{2}w) + f(r) \, .
}
Then $|w|\leq 1$, \eqref{existC18}, \eqref{existC19}, \eqref{existC21.2}, and
\eqref{existC22} imply that
$(v-u)''(r_{m}) > 0$. Therefore there exists an $\sigma > 0$ such that
$v'(r) > u'(r)$ for $r_{m}\leq r < r_{m}+\sigma$ and hence
$v(r) > u(r)$ for  $r_{m}\leq r < r_{m}+\sigma$. Let $r_{*}$ be the first
$r$ greater than $r_{m}$ for which $v'(r)=u'(r)$. Using 
$v(r_{*})\geq u(r_{*})$, $|w|\leq 1$, \eqref{existC18}, \eqref{existC19}, and
\eqref{existC22}, we see that $(u-v)''(r_{*}) > 0$ which contradicts
$v'(r_{*})=u'(r_{*})$. Therefore $v'(r) > u'(r)$ for all $r\geq r_{m}$ which
implies that
\leqn{existC23}{
 1-w(r)^{2} < v(r) \quad \forall \; r\geq r_{m} \, .
} 

The general solution to \eqref{existC21.1} is
\leqn{existC24}{
v = \frac{C_{1}}{r} + C_{2}r^{2} -\frac{1}{r}\int_{r_{m}}^{r}\rho^{1/2}w'(\rho)d\rho\, .
}
where $C_{1}$ and $C_{2}$ are arbitrary constants. So then
\eqn{existC25}{
C_{2} = \frac{1}{3r}\left(\frac{v}{r}+ v'+\frac{w'}{r^{1/2}}\right) 
}
and hence
\alin{existC26}{
C_{2} & = \frac{1}{3r_{m}}\left( \frac{u(r_{m})}{r_{m}} +
u'(r_{m}) + \frac{w'(r_{m})}{r_{m}^{1/2}}\right) && \text{by \eqref{existC21.2}}\\
& = 0  && \text{by \eqref{existC20}} \, .
}
Therefore
\leqn{existC27}{
1-w^{2}(r) \leq \frac{C_{1}}{r} - \frac{1}{r}\int_{0}^{r}\rho^{1/2}w'(\rho)d\rho 
\quad \forall \; r\geq r_{m} \, .
}
As $w'= \text{o}(r^{-1})$ it is easy to see that there exists a $C > 0$ such
that
\eqn{existC28}{
1-w^{2}(r) \leq \frac{C}{r^{1/2}} \quad \forall \; r\geq r_{m} \, .
}
Using the same arguments as in lemma \ref{existD} it follows from
this inequality that $w'= \text{O}(r^{-3/2})$. Using this
back in \eqref{existC27} we see that 
\eqn{existC28a}{
1-w^{2}(r) \leq \frac{C}{r^{\epsilon}} \quad \forall \; r\geq r_{m} \, 
}
for any $\epsilon \in (0,1)$. Again using the arguments from
lemma \ref{existD} we get that $w'= \text{O}(r^{-2\epsilon})$
for any $\epsilon \in (0,1)$.
\end{proof}

Since $\lim_{r\rightarrow \infty}w(r) = 1$, we have
$1-w(r) = \int_{r}^{\infty} w'(\rho) d\rho$
and hence
\leqn{existC30}{
|1-w(r)| \leq \int_{r}^{\infty} |w'(\rho)|d\rho \, .
}
But $w'= \text{O}(r^{-2\epsilon})$ by lemma \ref{existF}, and hence 
$1-w(r) = \text{O}(r^{-2\epsilon+1})$ by \eqref{existC30}. Writing \eqref{YMd.3.1} as
\leqn{existC31}{
w'' = -\alpha' w' + \frac{(w-1)(w+1)w}{r^{2}}
}
we see that $w''= \text{O}(r^{-2\epsilon-1})$ since $|w|\leq 1$, $w'= \text{O}(r^{-2\epsilon})$,
$1-w(r) = \text{O}(r^{-2\epsilon+1})$, and $\alpha' = \text{O}(r^{-2})$.
Using \eqref{YMd.3.2}, $\lim_{r\rightarrow\infty}\alpha(r)=\infty$, and similar
arguments, it is straightforward to show that $\alpha = \text{O}(r^{-1})$
and $\alpha'' = \text{O}(r^{-2})$.
\end{proof}

We can now use the previous proposition to show that the gauge potential
and its corresponding field arising from the solutions in
theorem \ref{existA} lie in certain weighted spaces. 
\begin{prop} \label{smoothAA}\mnote{[smoothAA]}
Suppose $(w(r),\alpha(r))$ is one of the solutions from theorem
\ref{existA}.
If $W_{\alpha}$ is given by \eqref{Wsplit} for $\alpha = 1,2,3$,
$W_{0}=0$, and
\leqn{FW}{
\quad F^{W}_{\alpha\beta} = \partial_{\alpha}W_{\beta} - \partial_{\alpha}
W_{\beta} + [W_{\alpha},W_{\beta}] \, 
}
then 
$W_{\alpha} \in \Ac^{2,p}_{\delta_{1}}$ for
any $\delta_{1} > -1$, $1<p<\infty$ and $F^{W}_{\alpha\beta} \in
\W^{2,p}_{\delta_{2}}(\Rbb^{3},\sU{2})$ for any $\delta_{2} > -2$, 
$1<p<\infty$.
\end{prop}
\begin{proof}
A short calculation shows that non-zero components of $F^{W}_{\alpha\beta}$
are
\leqn{FWa}{
F^{W}_{ij} = \epsilon_{ijk}\left[ \frac{w'(r)}{r}\left(\delta^{kl} -
\frac{x^{k}x^{l}}{r^2}\right)
+ \frac{w^2-1}{r^{4}} x^{k}x^{l}\right]\tau_{l}  \, .
}
The proof then follows directly from theorem 
\ref{existA}, proposition \ref{existC},
formulas \eqref{Wsplit}, \eqref{FWa}, and the definition
of the spaces $\Ac^{2,p}_{\delta_{1}}$,
$\W^{2,p}_{\delta_{2}}(\Rbb^{3},\sU{2}^3)$.
\end{proof}

\begin{rem} \label{newtrem} \mnote{[newtrem]}
For the remainder of this report we will always assume that
$(w(r),\alpha(r))$ is one of the solutions to the Euclidean 
Yang-Mills-dilaton equations \eqref{YMd.3.1}-\eqref{YMd.3.2}
from theorem \eqref{existA}.
\end{rem}

\subsect{slYMd}{Solutions of the linearized Yang-Mills equations}

As will be seen later in section \ref{ymdsol}, the main obstacle
to having a complete proof of the existence of EYMd solutions is
that we do not yet have a complete understanding of
the solutions to the lYMd equations
\lgath{isoB5}{
v'' + \phi'w'+\alpha'v'-\frac{(3w^{2}-1)}{r^{2}}v = 0\, , \label{isoB5.1} \\
(r^{2}\phi')'-e^{\alpha}\left({w'}^{2}+\frac{(w^{2}-1)^{2}}{2 r^{2}}\right)
\phi - 2e^{\alpha}\left(w'v'+\frac{(w^{2}-1)}{r^{2}}wv\right)=0 \, ,
\label{isoB5.2}
}
that satisfy  the boundary conditions 
\leqn{bound}{
v(r) = \Ord(r^2) \quad \text{and} \quad \phi(r) = \Ord(1) \quad
\text{as $r\rightarrow 0$.} 
}
Using the fact that $w(r)$ and $\alpha(r)$ are analytic in a neighborhood
and that $w(r) = 1 -\beta r^2 + \Ord(r^4)$ and $\alpha(r) = \Ord(1)$
as $r\rightarrow 0$ (see remark \ref{existAA}) , it can be shown
using theorem  5.0.6 of  \cite{Tthesis} that there are exactly  
two $C^2$ linearly independent solutions
$(v_{1}(r),\phi_{1}(r))$ and $(v_{2}(r),\phi_{2}(r))$ to \eqref{isoB5.1}-
\eqref{isoB5.2}
which satisfy the boundary conditions \eqref{bound}. The solutions
are uniquely determined by the their expansions near $r=0$:
\leqn{lYMdexp1}{
v_{1}(r) = -\beta r^2 + \Ord(r^4)\, \quad  \phi_{1}(r) = -1 + \Ord(r^2)\,
}
and
\leqn{lYMdexp2}{
v_{2}(r) = -r^2 + \Ord(r^4) \quad \phi_{2}(r) = \Ord(r^2) \, 
}
as $r\rightarrow 0$.
It also
follows from theorem 5.0.6  of \cite{Tthesis} that the solutions  
are analytic in a neighborhood of $r=0$. This coupled with the fact that
$(w(r),\alpha(r))$ are analytic for $r > 0$ implies that
the solutions $(v_{1}(r),\phi_{1}(r))$ and $(v_{2}(r),\phi_{2}(r))$
are also analytic for $r>0$. 

The following lemma shows that we can exactly determine 
the solution $(v_{1}(r),\phi_{1}(r))$.
\begin{lem} \label{isoD} \mnote{[isoD]}
\leqn{isoD1}{
\phi_{1}(r) = \frac{r}{2} \alpha'(r) -1\quad v_{1}(r) = \frac{r}{2} w'(r)
}
\end{lem}
\begin{proof}
From \eqref{isoB6} we see that 
$w_{\beta}(r) = w(e^{\beta/2}r)$ and 
 $\alpha_{\beta}=\alpha(e^{\beta/2}r)-\beta$ defines a 
1-parameter family of solutions passing through the
solution $(w(r),\alpha(r))$. Therefore,
\eqn{isoD2}{
v(r) := \frac{d\,}{d\beta}\Bigl|_{\beta=0}w_{\beta}(r)
= \frac{r}{2}w'(r)
\quad \text{and} \quad
\phi(r) := \frac{d\,}{d\beta}\Bigl|_{\beta=0}\alpha_{\beta}(r)
=\frac{r}{2}\alpha'(r)-1
}
must satisfy the linearized equations \eqref{isoB5.1}-\eqref{isoB5.2}.
The fact that this solution 
satisfies \eqref{lYMdexp1} follows from the expansions
$w(r) = 1-\beta r^2 + \Ord(r^4)$ and $\alpha(r) = \Ord(1)$.
\end{proof}
The fall of conditions for $w(r)$ and $\alpha(r)$
as $r\rightarrow \infty$ imply that
\leqn{asymlin1}{
\lim_{r\rightarrow \infty} (v_{1}(r),\phi_{1}(r)) = (0,-1) \, .
}
At the present, we do not have a understanding of the asymptotic behavior
of $r\rightarrow \infty$ for the solution $(v_{2}(r),\phi_{2}(r))$.
This is the main obstacle in our having a complete existence proof. 
However, we \emph{conjecture} that
\leqn{asymlin2}{
\lim_{r\rightarrow \infty} |v_{2}(r)| + |\phi_{2}(r)| = \infty \, .
}
If this were not true, then there would exist a bounded solution
to the lYMd equations. It would then be natural to expect
that there exists a 1-parameter family of bounded solutions to
the YMd equations which when differentiated gives
rise to the bounded lYMd solution.
As shown above, this is how the 
solution $(v_{1}(r),\phi_{1}(r))$ arises.
We note, however, that numerical evidence does not support the existence of
1-parameter families of bounded solutions that pass through
the YMd solutions from theorem \ref{existA} other that
the family that arises via scaling \eqref{isoB6}. These
solutions appear to be unique up to scaling. 

The difficulty in proving \eqref{asymlin2} is that even though
\eqref{isoB5.1}-\eqref{isoB5.2} are linear equations,
we do not have very much information about the coefficients
because they depend on the functions $w(r)$ and $\phi(r)$
of which we know very little. This makes it difficult
to determine the behavior of the solution $(v_{2}(r),\phi_{2}(r))$.
However, we will show in the following sections how to prove existence of 
solutions to the EYMd equations under the assumption that 
\eqref{asymlin2} is true.

%% file: smooth.tex
\sect{smooth}{Differentiability of the field equations}

In this section we establish that the reduced
field equations and the YMd equations define
differentiable maps. In fact they define analytic
maps. Before we proceed we first introduce some definitions.

Let $\mathcal{L}_{k}(B_{1},B_{2})$ denote the Banach space
of $k$-linear and continuous maps from the Banach space
$B_{1}$ into $B_{2}$ with norm
\eqn{multinorm}{
\norm{T}_{\mathcal{L}_{k}(B_{1},B_{2})} :=
\sup\{\,\|T(x_{1},\ldots,x_{k})\|_{B_{2}}\,|\, \sup\{\|x_{1}\|_{B_{1}},
\ldots,\|x_{k}\|_{B_{1}}\} \leq 1\, \} \, .
}

\begin{Def} \label{analyticA} \mnote{[analyticA]}
Let $X_{1}$ and $X_{2}$ be Banach spaces and let 
$U$ be an open subset of $X_{1}$. 
A map $T : U \rightarrow V_{2}$ is said to be \emph{analytic}
if for each $x\in U$ there exists a $R > 0$ and
a sequence $T_{k} \in \mathcal{L}_k(X_{1},X_{2})$ of
$k$-linear symmetric maps such that 
\eqn{analyticA1}{
\sum_{k=0}^{\infty} R^{k}\norm{T_{k}}_{\mathcal{L}_{k}(X_{1},X_{2})} < 
\infty   
}
and
\eqn{analyticA2}{
T(y) = \sum_{k=0}^{\infty} T_{k}(y-x,\ldots,y-x) \quad 
\text{for all $y$ with $\norm{y-x}_{X_{1}}< R$.}
}
We use $\Cw(U,X_{2})$ to denote the set of all
the analytic maps from $U$ to $X_{2}$.
\end{Def}
An open ball in a Banach space $X$ will be denoted by
\eqn{ball}{
B_{X}(x;R) := \{\, y \in X \, |\, \|x-y\|_{X} < R \} 
}
We then have the following useful proposition:
\begin{prop} \label{analyticB} \mnote{[analyticB]}
Suppose $u\in \Cw(B_{\Rbb^{n}}(0;R),\Rbb)$ satisfies
$u(0)=0$. Furthermore, suppose $X$ is a commutative
Banach algebra where $C$ is any constant such that
$\|x y\|_{X} \leq C\|x\|_{X}\|y\|_{X}$ for
all $x,y\in X$. Then the map
\eqn{analyticB2}{
\hat{u} : X^{n} \rightarrow X : (x_{1},\ldots,x_{n}) \mapsto
\sum_{|I|=1}^{\infty} \frac{1}{I!}\left(\partial^{I}u(0)\right)
x_{1}^{I_{1}}\ldots x_{n}^{I_{n}}
}
is of class $\Cw(B_{X}(0;\rho)^{n},X)$ for $\rho = R/C$.
\end{prop} 

Note that 
\leqn{gflatupA}{
(\underset{o}{\gb}{}^{\alpha\beta}) = \begin{pmatrix}
-\lambda & 0 & 0 & 0 \\
 0   & 1 & 0 & 0\\
 0 & 0 & 1 & 0 \\
 0 & 0 & 0 & 1
\end{pmatrix}
}
so that 
\leqn{gflatupB}{
(\underset{o}{\gb}{}^{\alpha\beta})\big|_{\lambda=0} = \begin{pmatrix}
 0 &  0 & 0 & 0 \\
 0   & 1 & 0 & 0\\
 0 & 0 & 1 & 0 \\
 0 & 0 & 0 & 1
\end{pmatrix}
}
As in \cite{Heil95}, we define for any weakly differentiable map
$u$
\eqn{raisediff}{
u^{,\alpha} := \underset{o}{\gb}{}^{\alpha\beta}\big|_{\lambda=0} u_{,\beta}
= \left\{ \begin{array}{cl} 
\partial_{\alpha}u & \text{for $\alpha \neq 0$} \\
0 & \text{for $\alpha = 0$}
\end{array}\right. \, .
}
We now collect some results from \cite{Heil95} concerning
the analyticity of various quantities  involving
the density $\U$. 

\begin{prop}{\emph{[Proposition 3.10,\cite{Heil95}]}}
 \label{HeilA}\mnote{[HeilA]} 
Suppose $p > 3/2$ and $-1<\delta <0$. 
Then for any $R >0$ there exists a $\Lambda > 0$ such
that the following maps are of class $\Cw$:
\gath{HeilA1}{
(-\Lambda,\Lambda) \times B_{\W^{2,p}_{\delta}(\Rbb^{3},\Sbb)}(0;R)
\rightarrow \W^{2,p}_{\delta}(\Rbb^{3},\Sbb) \; :\; 
(\lambda,\U) \mapsto (\gb^{\alpha\beta} - \underset{o}{\gb}{}^{\alpha\beta}) \\
(-\Lambda,\Lambda) \times B_{\W^{2,p}_{\delta}(\Rbb^{3},\Sbb)}(0;R)
\rightarrow \W^{2,p}_{\delta}(\Rbb^{3},\Sbb) \; :\; 
(\lambda,\U) \mapsto (\gb_{\alpha\beta} - \underset{o}{\gb}{}_{\alpha\beta}) \\
\intertext{and}
(-\Lambda,\Lambda) \times B_{\W^{2,p}_{\delta}(\Rbb^{3},\Sbb)}(0;R)
\rightarrow \W^{2,p}_{\delta}(\Rbb^{3}) \; :\; 
(\lambda,\U) \mapsto |\df|^{q/2}-1 
}
for $q=-3,-2,-1,1,2$. Moreover, the following expansions
are valid 
\gath{HeilA2}{
|\df|-1 = -4\lambda \U^{00} + \Ord(\lambda^{2}) \, ,\quad
\sqrt{\df}-1 = -2\lambda \U^{00} + \Ord(\lambda^{2}) \, ,\\
\frac{1}{\sqrt{\df}}-1 = 2\lambda \U^{00} + \Ord(\lambda^{2}) \, , \quad
(\gb_{\alpha\beta} - \underset{o}{\gb}{}_{\alpha\beta}) = -4\lambda 
(\delta^{0}_{\alpha}\delta^{0}_{\beta})\U^{00} + \Ord(\lambda^{2}) \, .
}
\end{prop}

\begin{prop}{\emph{[Proposition 6.2,\cite{Heil95}]}} 
\label{HeilB}\mnote{[HeilB]}
Suppose $p > 3$ and $-1 < \delta < 0$. Then for any $R>0$ there
exists a $\Lambda >0$ such that the Christoffel symbols
\eqn{HeilB1}{
\Gamma^{\alpha}_{\beta\gamma} : 
(-\Lambda,\Lambda) \times B_{\W^{2,p}_{\delta}(\Rbb^{3},\Sbb)}(0;R)
\rightarrow \W^{1,p}_{\delta-1}(\Rbb^{3}) 
} 
are of class $\Cw$ for all $\alpha,\beta,\gamma = 0,1,2,3$. Moreover,
the following expansion is valid
\eqn{HeilB2}{
\Gamma^{\alpha}_{\beta\gamma} = \Gamma^{\alpha}_{\beta\gamma}\big|_{\lambda = 0}
+ \Ord(\lambda)
}
where
\eqn{HeilB3}{
\Gamma^{\alpha}_{\beta\gamma}\big|_{\lambda = 0} = \left\{
\begin{array}{ll} \U^{00}{}_{,\alpha} & \text{if $\beta=\gamma=0$ and
$\alpha \neq 0$}\\
0 & \text{otherwise} 
\end{array} \right. \, .
}
\end{prop}

\begin{prop} \label{HeilD}\mnote{[HeilD]}
Suppose $p > 3$ and $-1 < \delta < 0$. Then for any
$R>0$ there exists a $\Lambda$ such that the map
\eqn{HeilD1}{
(E-\Delta) :(-\Lambda,\Lambda) \times B_{\W^{2,p}_{\delta}(\Rbb^{3},\Sbb)}(0;R)
\rightarrow \W^{0,p}_{\delta-2}(\Rbb^{3},\Sbb)
 \; :\;
(\lambda,\U) \mapsto (E^{\alpha \beta} - \Delta\U^{\alpha\beta})
}
is of class $\Cw$ where $E^{\alpha\beta}$ is defined by
\eqref{reduceda1}. Moreover,
\eqn{HeilD2}{
\D_{2}(E-\Delta)(0,\U)\cdot\delta\U =
(\delta\U^{00,\alpha}\U^{00,\beta}
+\delta\U^{00,\beta}\U^{00,\alpha} - 
\underset{o}{\gb}{}^{\alpha\beta}\big|_{\lambda=0} \delta \U^{00,\gamma}
\U^{00}{}_{,\gamma}) \, .
}
\end{prop}
\begin{proof}
The proof of this proposition is contained in the proof
of proposition 4.2 in \cite{Heil95}.
\end{proof}

Let 
\leqn{YMdB1}{
\Upsilon^{2} := \gb^{\alpha\beta}\left(\psi_{,\alpha\beta} - \Gamma^{\mu}_{\alpha\beta}
\psi_{,\mu} - \frac{\kappa\ell_{Y}}{\ell_{d}}\frac{e^{2\kappa \psi}}{\sqrt{|\df|
}}
\gb^{\mu\nu}\ip{F_{\alpha\mu}}{F_{\beta\nu}}\right) \, ,
}
and
\leqn{YMdB2}{
\Upsilon^{1}_{ j} := \gb^{\alpha\nu}\left(F_{\alpha j,\nu} - \Gamma^{\mu}_{\alpha\nu}
F_{\mu j} - \Gamma^{\mu}_{j\nu} F_{\alpha\mu} +
2\kappa \psi_{,\nu} F_{\alpha j} + [A_{\nu},F_{\alpha   j}]\right)
\, .
}
The YMd equations are then
$\Upsilon^{1} = (\Upsilon^{1}_{j}) = 0$ and $\Upsilon^{2} = 0$.
Note that the Yang-Mills equation $\Upsilon^{1}_{j} = 0$
appears to be missing a component. However, due to our assumption
that the fields a static and spherically symmetric it follows
that $F_{\alpha 0} =0$ and $\Gamma^{k}_{l0}=0$ and hence
that $\beta =0$ component of the Yang-Mills equation \eqref{Ymda1}
is automatically satisfied.

We will split the gauge potential $A$ and the dilaton field $\psi$ 
as follows
\leqn{gsplit}{
 A(x) = W(r) + Y(x) = W_{\alpha}(r)dx^{\alpha} + Y_{\alpha}(x)dx^{\alpha} 
}
and
\leqn{dsplit}{
\psi(x) = \alpha(r) + \xi(x)
}
where $W_{\alpha}(r) = \delta_{\alpha}^{i}r^{-2}(w(r) -1)
\epsilon_{i}{}^{j}{}_{k}x^{k}\tau_{j}$ and 
and $\alpha(r)$ are to be considered as \emph{fixed}.
Recall that we are assuming that $(w(r),\alpha(r))$
is one of the solutions to the Euclidean Yang-Mills-dilaton equations
from theorem \ref{existA}.
Under the splitting \eqref{gsplit}, the gauge potential 
decomposes as
\leqn{Fsplit}{
F_{\alpha\beta} = F^{W}_{\alpha\beta} + F^{Y}_{\alpha\beta}
+ [Y_{\alpha},W_{\beta}] + [W_{\alpha},Y_{\beta}]
} 
where $F^{W}$ is defined by \eqref{FW} 
and
\leqn{FY}{
F^{Y}_{\alpha\beta} := \partial_{\alpha}Y_{\beta} - \partial_{\beta}
Y_{\alpha} + [Y_{\alpha},Y_{\beta}] \; .
}
Note that only $F_{\alpha\beta}$ and $F^{W}_{\alpha\beta}$
define field strengths. The quantity $F^{Y}_{\alpha\beta}$ does not define
a field strength as $Y_{\alpha}$ does not transform as a gauge potential
under gauge transformations. 
\begin{prop} \label{smoothAC} \mnote{[smoothAC]}
Suppose $-2 > \delta > -1$ and $p > 3$. Then for any $R > 0$ 
and $\alpha,\beta, \gamma = 0,1,2,3$ the following maps are $\emph{\text{C}}^{\omega}$
\eqn{smoothAC1}{
\text{B}_{\W^{2,p}_{\delta}(\Rbb^{3},\sU{2}^{3})}(0;R) \rightarrow
\W^{1,p}_{\delta-1}(\Rbb^{3},\sU{2})\; :\; (Y_{j}) \mapsto F_{\alpha\beta} \, ,
}
and
\eqn{smoothAC2}{
\text{B}_{\W^{2,p}_{\delta}(\Rbb^{3},\sU{2}^{3})}(0;R) \rightarrow
\W^{0,p}_{\delta-2}(\Rbb^{3},\sU{2})\; :\; (Y_{j}) \mapsto [A_{\alpha},F_{\beta\gamma}]\, , 
}
where $A_{\alpha}$ and $F_{\alpha\beta}$ are given by the formula \eqref{gsplit} and \eqref{Fsplit},
respectively.
\end{prop}
\begin{proof}
The proof is a direct consequence of lemma \ref{multiply} and proposition \ref{smoothAA}.
\end{proof}
\begin{prop} \label{smoothA}\mnote{[smoothA]}
Suppose $p > 3$ and $-1<\delta_{1} <0$ and
$-2< \delta_{2} < -1$. 
Then for any $R >0$ there exists a $\Lambda > 0$ such
that the map
\gath{smoothA1}{
\Upsilon : (-\Lambda,\Lambda) \times 
B_{\W^{2,p}_{\delta_{1}}(\Rbb^{3},\Sbb)}(0;R) \times
B_{\W^{2,p}_{\delta_{2}}(\Rbb^{3},\sU{2}^{3})}(0;R) \times
B_{\W^{2,p}_{\delta_{1}}(\Rbb^{3})}(0;R) \\
\longrightarrow \W^{0,p}_{\delta_{1}-2}(\Rbb^{3})\times
\W^{0,p}_{\delta_{2}-2}(\Rbb^{3},\sU{2}^{3}) \; :\;
(\lambda,\U,Y,\xi) \longmapsto (\Upsilon^{1},\Upsilon^{2})
}
is of class $\Cw$. 
\end{prop}
\begin{proof}
Follows easily from lemma \ref{multiply} and  propositions
\ref{analyticB}, \ref{HeilA}, \ref{HeilB}, and \ref{smoothAC}.
\end{proof}

\begin{prop} \label{smoothB}\mnote{[smoothB]}
Suppose $p > 3$ and $-1<\delta_{1} <0$ and
$-2< \delta_{2} < -1$.
Then for any $R >0$ there exists $\Lambda > 0$ and
$\epsilon > 0$ such
that the maps
\gath{smoothB1}{
T : (-\Lambda,\Lambda) \times
B_{\W^{2,p}_{\delta_{1}}(\Rbb^{3},\Sbb)}(0;R) \times
B_{\W^{2,p}_{\delta_{2}}(\Rbb^{3},\sU{2}^{3})}(0;R) \times
B_{\W^{2,p}_{\delta_{1}}(\Rbb^{3})}(0;R) \\
\longrightarrow \W^{1,p}_{\delta_{1}-(2+\epsilon)}(\Rbb^{3},\Sbb)
\; :\;
(\lambda,\U,Y,\xi) \longmapsto (T^{\alpha\beta})
}
and
\gath{smoothB2}{
\Tc : (-\Lambda,\Lambda) \times
B_{\W^{2,p}_{\delta_{1}}(\Rbb^{3},\Sbb)}(0;R) \times
B_{\W^{2,p}_{\delta_{2}}(\Rbb^{3},\sU{2}^{3})}(0;R) \times
B_{\W^{2,p}_{\delta_{1}}(\Rbb^{3})}(0;R) \\
\longrightarrow \W^{1,p}_{\delta_{1}-(2+\epsilon) }(\Rbb^{3},\Sbb)
\; :\;
(\lambda,\U,Y,\xi) \longmapsto (\Tc^{\alpha\beta})
}
are of class $\Cw$.
\end{prop}
\begin{proof}
Follows easily from lemma \ref{multiply} and  propositions
\ref{analyticB}, \ref{HeilA}, and \ref{smoothAC}. 
\end{proof}

We now prove spherically symmetric versions of
propositions \ref{HeilD}, \ref{smoothA} and  \ref{smoothB}.

\begin{prop} \label{smoothC}\mnote{[smoothC]}
Suppose $p > 3$ and $-1 < \delta < 0$. Then for any
$R>0$ there exists a $\Lambda$ such that the map
\eqn{smoothC1}{
(E-\Delta) :(-\Lambda,\Lambda) \times B_{\Uc^{2,p}_{\delta}}(0;R)
\rightarrow \U^{0,p}_{\delta-2}
 \; :\;
(\lambda,\U) \mapsto (E^{\alpha \beta} - \Delta\U^{\alpha\beta})
}
is of class $\Cw$. Moreover,
\eqn{smoothC2}{
\D_{2}(E-\Delta)(\lambda,\U)\cdot\delta\U =
(\delta\U^{00,\alpha}\U^{00,\beta}
+\delta\U^{00,\beta}\U^{00,\alpha} - 
\underset{o}{\gb}{}^{\alpha\beta}\big|_{\lambda=0} \delta \U^{00,\gamma}
\U^{00}{}_{,\gamma}) \, .
}
\end{prop}
\begin{proof}
Given $R$, let $\Lambda$ be determined as in proposition
\ref{HeilD}. By straightforward calculation it can be shown that
if $\U \in \sCo(\Rbb^{3},\Sbb)\cap B_{\W^{2,p}_{\delta}(\Rbb^{3},\Sbb)}(0;R)$
then \\$(E - \Delta)(\U)$ $\in \sCo(\Rbb^{3},\Sbb)$. Consequently
$(E-\Delta)$ $\left(\sCo(\Rbb^{3},\Sbb)\cap B_{\W^{2,p}_{\delta}(\Rbb^{3},\Sbb)}(0;R)\right)$
$\subset \sCo(\Rbb^{3},\Sbb)$.
Therefore 
$(E-\Delta)\left(B_{\Uc^{2,p}_{\delta}}(0;R)\right)\subset \Uc^{0,p}_{\delta-2}$
by the density of $\sCo(\Rbb^{3},\Sbb)$ in $\Uc^{2,p}_{\eta}$ for $\eta \in \Rbb$, 
and continuity of the map $(E-\Delta)$ by proposition \ref{HeilD}.
The proposition now follows from proposition \ref{HeilD}.
\end{proof}

\begin{prop} \label{smoothD}\mnote{[smoothD]}
Suppose $p > 3$ and $-1<\delta_{1} <0$ and
$-2< \delta_{2} < -1$. 
Then for any $R >0$ there exists a $\Lambda > 0$ such
that the map
\gath{smoothD1}{
\Upsilon : (-\Lambda,\Lambda) \times 
B_{\Uc^{2,p}_{\delta_{1}}}(0;R) \times
B_{\Ac^{2,p}_{\delta_{2}}}(0;R) \times
B_{\Dc^{2,p}_{\delta_{1}}}(0;R) \\
\longrightarrow \Ac^{0,p}_{\delta_{2}-2}\times
\Dc^{0,p}_{\delta_{1}-2}
\; :\;
(\lambda,\U,Y,\xi) \longmapsto (\Upsilon^{1},\Upsilon^{2})
}
is of class $\Cw$.
\end{prop}
\begin{proof}
As in the proof of proposition \ref{smoothC}, straightforward
calculation shows that  if $\U \in \sCo(\Rbb^{3},\Sbb)\cap 
B_{\W^{2,p}_{\delta}(\Rbb^{3},\Sbb)}(0;R)$, $Y \in \sAo\cap 
B_{\W^{2,p}_{\delta_{2}}(\Rbb^{3},\sU{2}^{3})}(0;R)$ and 
$\xi \in \sCo(\Rbb^{3})\cap B_{\W^{2,p}_{\delta_{1}}(\Rbb^{3})}(0;R) $
then $\Upsilon(Y,\alpha) \in \sCo(\Rbb^{3})\cap \text{C}^{2}\times\sAo\cap \text{C}^{2}$.
We then argue in the same manner as proposition \ref{smoothC}.
\end{proof}

\begin{prop} \label{smoothE}\mnote{[smoothE]}
Suppose $p > 3$ and $-1<\delta_{1} <0$ and
$-2< \delta_{2} < -1$.
Then for any $R >0$ there exists $\Lambda > 0$ and $\epsilon > 0$ such
that the maps
\gath{smoothE1}{
T : (-\Lambda,\Lambda) \times
B_{\Uc^{2,p}_{\delta_{1}}}(0;R) \times
B_{\Ac^{2,p}_{\delta_{2}}}(0;R) \times
B_{\Dc^{2,p}_{\delta_{1}}}(0;R) \\
\longrightarrow \Uc^{1,p}_{\delta_{1}-(2+\epsilon)}
\; :\;
(\lambda,\U,Y,\xi) \longmapsto (T^{\alpha\beta})
}
and
\gath{smoothE2}{
\Tc : (-\Lambda,\Lambda) \times
B_{\Uc^{2,p}_{\delta_{1}}}(0;R) \times
B_{\Ac^{2,p}_{\delta_{2}}}(0;R) \times
B_{\Dc^{2,p}_{\delta_{1}}}(0;R) \\
\longrightarrow \Uc^{1,p}_{\delta_{1}-(2+\epsilon)}
\; :\;
(\lambda,\U,Y,\xi) \longmapsto (\Tc^{\alpha\beta})
}
are of class $\Cw$.
\end{prop}
\begin{proof}
See the proofs of propositions \ref{smoothC} and \ref{smoothD}.
\end{proof}

From \eqref{embedA} and propositions \ref{slaplaceB},
\ref{smoothC} and \ref{smoothE} we get the following:

\begin{prop} \label{smoothF} \mnote{[smoothF]}
Suppose $-1<\delta_{1}<0$, $-2<\delta_{2}<-1$ and $p > 3$. Then
for any $R>0$ there exists a $\Lambda >0$
such that
\gath{smoothF}{
\Xi : (-\Lambda,\Lambda) \times
B_{\Uc^{2,p}_{\delta_{1}}}(0;R) \times
B_{\Ac^{2,p}_{\delta_{2}}}(0;R) \times
B_{\Dc^{2,p}_{\delta_{1}}}(0;R) 
\longrightarrow \Uc^{2,p}_{\delta{1}} \\\; : \;
(\lambda,\U,Y,\xi) \mapsto \left(\U^{\alpha\beta} - \Delta^{-1}\left\{
\Tc^{\alpha\beta} -(E^{\alpha\beta}-\Delta\U^{\alpha\beta})\right\}\right)
}
is of class $\Cw$.
\end{prop}
From the definition of $\Xi$ it is clear that the reduced
field equations \eqref{reduced2} are equivalent to 
$\Xi = 0$ .

%% file: reduced.tex
\sect{redsol}{Solving the reduced field equations}

We now employ the same method as in \cite{Heil95} to find
solutions to the reduced field equations. Namely,
we first solve the reduced equations for $\lambda = 0$,
and then use an implicit function argument to show that there
exist a solution for $\lambda$ small enough.

\subsect{lo}{$\lambda = 0$}

Assume $-1<\delta_{1} <0$ , $-2<\delta_{2}<-1$, $p > 3$ and for
fixed $R>0$ let $\Lambda >0$ be as in proposition \ref{smoothF}.
From the expansions in proposition \ref{HeilA} and 
\eqref{einsta3} we see that
\leqn{Eo}{
E^{\alpha\beta}\big|_{\lambda=0} = \Delta \U^{\alpha\beta} + \left\{\begin{array}{cl}
-\U^{00,\alpha}\U^{00,\beta} + \Half \delta^{\alpha\beta}
|\grad \U^{00}|^{2} & \text{if $\alpha \neq 0, \beta \neq 0$} \\
0 & \text{otherwise} \end{array} \right.,
}
and
\alin{strengo}{
\Tc^{\alpha \beta}\big|_{\lambda=0} =& 2\pi G \ell_{d} \left( \left(\underset{o}{\gb}{}^{\alpha\mu}
\underset{o}{\gb}{}^{\beta\nu}\right)\bigg|_{\lambda=0}
\psi_{,\mu}\psi_{,\nu} - \Half \left(\underset{o}{\gb}{}^{\alpha\beta}
\underset{o}{\gb}{}^{\mu\nu}\right)\bigg|_{\lambda=0}\psi_{,\mu}\psi_{,\nu}
\right) + \notag \\
& 4\pi G \ell_{Y} e^{2\kappa\psi}\left( \left(\underset{o}{\gb}{}^{\alpha\mu}
\underset{o}{\gb}{}^{\beta\nu}\underset{o}{\gb}^{\sigma\tau}\right)\bigg|_{\lambda=0}
\ip{F_{\mu\sigma}}{F_{\nu\tau}}-\Quarter \left(\underset{o}{\gb}{}^{\mu\nu}
\underset{o}{\gb}^{\sigma\tau}\underset{o}{\gb}{}^{\alpha\beta}\right)\bigg|_{\lambda=0}
\ip{F_{\mu\sigma}}{F_{\nu\tau}} \right) \label{streng1} \, .
}
Therefore
\eqn{Ta}{
\Tc^{00}\big|_{\lambda=0} = 0\,, \quad \Tc^{0j}\big|_{\lambda=0} = 0\, , \quad \text{and} \quad
\Tc^{ij}\big|_{\lambda=0} = 4\pi G\Tn^{ij}\, ,
}
where
\alin{Tb}{
\Tn^{ij} :=& \Half\ell_{d} \left(\delta^{ik}
\delta^{jl} \psi_{,k}\psi_{,l} - \Half \delta^{ij}\delta^{kl}
\psi_{,k}\psi_{,l} \right) + \notag \\
& \ell_{Y} e^{2\kappa\psi}\left( \delta^{ik}\delta^{lm}\delta^{jn}
\ip{F_{kl}}{F_{mn}}-\Quarter \delta^{ij}\delta^{kl}\delta^{mn}
\ip{F_{km}}{F_{ln}} \right) 
}
is the stress-energy tensor for the Euclidean YMd equations
on $\Rbb^3$.
So then
\eqn{solAa}{
\Xi(0,\U,Y,\xi) =  0 \quad \Longleftrightarrow 
\left\{ \begin{array}{l}
\Delta \U^{00} = 0 \, \\
\Delta \U^{0j} = 0 \, \\
\Delta \U^{ij} = \U^{00,i}\U^{00,j} - \Half\delta^{ij}
|\text{grad}\,\U^{00}|^2+4\pi G \Tn^{ij}
\end{array} 
\right.\, .
}
The first equation $\Delta \U^{00} =0$ can be interpreted
as the Newtonian gravitational equation
for the gravitational potential $\U^{00}$ \cite{Lott92}.
The vanishing of the mass density ($\Tc^{00}\big|_{\lambda=0}=0$)
decouples the Newtonian potential from the YMd fields in the
limit $\lambda \rightarrow 0$. For other matter fields such
as perfect fluids, this decoupling does not occur as
$\Tc^{00}\big|_{\lambda=0}\neq 0$ \cite{Lott92,Heil95}.
 
The invertibility of the Laplacian (Theorem \ref{laplace}) then
implies that
\leqn{solB}{
\U^{00} = 0\,,\quad \U^{0j} = 0\, ,\quad
\text{and} \quad \U^{ij} = 4\pi G \Delta^{-1}\Tn^{ij}
}
solve $\Xi(0,\U,Y,\xi)=0$ for any $Y \in B_{\Ac^{2,p}_{\delta_{2}}}(0;R)$ and
$\xi \in B_{\Dc^{2,p}_{\delta_{1}}}(0;R)$.

\subsect{lno}{$\lambda \neq 0$}

\begin{prop} \label{redsolA} \mnote{[redsolA]}
Suppose $-1<\delta_{1} < 0$, $-2<\delta_{2} < -1$, and
$p > 3$. Then  
there exists a $\Lambda > 0$, $\epsilon > 0$
and a $C^{\infty}$ map
\eqn{redsolA1}{
\Uh : (-\Lambda,\Lambda) \times B_{\Ac^{2,p}_{\delta_{2}}}(0;\epsilon)\times 
B_{\Dc^{2,p}_{\delta_{1}}}(0;\epsilon)\rightarrow \Uc^{2,p}_{\delta_{1}}
\; :\: (\lambda,Y,\xi) \rightarrow \Uh(\lambda,Y,\xi) = (
\Uh^{\alpha\beta}(\lambda,Y,\xi))
} such that
$\Xi(\lambda,\Uh(\lambda,Y,\xi),Y,\xi) = 0$
for all $(\lambda,Y,\xi)$ $\in  
(-\Lambda,\Lambda) \times B_{\Ac^{2,p}_{\delta_{2}}}(0;\epsilon)\times 
B_{\Dc^{2,p}_{\delta_{1}}}(0;\epsilon)$.
Moreover, $\Uh$ satisfies
$\Uh^{00}(0,0,0) = 0$, $\D_{2} \Uh^{00}(0,0,0) = 0$, and
$\D_{3} \Uh^{00}(0,0,0) = 0$.
\end{prop}
\begin{proof}
Fix $R > 0$ and let $\Lambda>0$ be chosen so that
the maps $\Xi$, $E-\Delta$ and  $\Tc$ are of class $\Cw$ which we can do by
propositions \ref{smoothC}, \ref{smoothE}, and \ref{smoothF}. Then
we can solve $\Xi(0,\U,0,0) = 0$
by \eqref{solB}. Let $\U_{b}$ denote the solution. Note that
$\U^{00}_{b}=0$ by \eqref{solB}. So
$\D_{2}(E-\Delta)(\lambda,\U_{b}) = 0$
by proposition \eqref{smoothC}. From the expansions in proposition \ref{HeilA},
and formula \eqref{einsta3} it follows that
$\D_{2}\Tc(0,\U_{b},0,0) = 0$.
Therefore from the definition of $\Xi$ it is clear that
\leqn{redsolA6}{
\D_{2}\Xi(0,\U_{b},0,0)= \id_{\Uc^{2,p}_{\delta_{1}}}\, ,
}
and hence by the implicit function theorem there exists a $\bar{\Lambda} > 0$,
$\epsilon > 0$, and a
$C^{\infty}$ map
\eqn{redsolA7}{
\Uh : (-\bar{\Lambda},\bar{\Lambda}) \times B_{\Ac^{2,p}_{\delta_{2}}}(0;\epsilon)\times 
B_{\Dc^{2,p}_{\delta_{1}}}(0;\epsilon) 
\rightarrow \Uc^{2,p}_{\delta_{1}}
\; :\: (\lambda,Y,\xi) \rightarrow \Uh(\lambda,Y,\xi) = (
\Uh^{\alpha\beta}(\lambda,Y,\xi))
}
such that
\leqn{redsolA8}{
\Xi(\lambda,\Uh(\lambda,Y,\xi),Y,\xi) = 0
}
for all $(\lambda,Y,\xi) \in
(-\bar{\Lambda},\bar{\Lambda}) \times B_{\Ac^{2,p}_{\delta_{2}}}(0;\epsilon)\times  
B_{\Dc^{2,p}_{\delta_{1}}}(0;\epsilon)$.
Differentiating \eqref{redsolA8} with respect to $Y$ and using
\eqref{redsolA6} we find
\leqn{redsolA9}{
\D_{2}\Uh^{00}(0,0,0) = - D_{3}\Xi^{00}(0,\U_{b},0,0)\, .
}
But 
\alin{redsolA10}{
D_{3}\Xi^{00}(\lambda,\U,Y,\xi)\cdot \delta Y& =  
\bigg( -4\pi G \Delta^{-1}\bigg\{ \frac{\ell_{Y}}{\sqrt{|\df|}} e^{-2\kappa 
\psi}\Big(
\gb^{\alpha\mu}\gb^{\beta\nu}\gb^{\sigma\tau} \Big[
\ip{\delta F_{\mu\sigma}}{F_{\nu\tau}}+ \\
&
\ip{F_{\mu\sigma}}{\delta F_{\nu\tau}}\Big]
-\Half \ip{\delta F_{\sigma\mu}}{F_{\nu\tau}} \gb^{\mu\nu}\gb^{\sigma\tau}
\gb^{\alpha\beta}\Big)
\bigg\} \bigg)
}
where 
\leqn{deltaF}{
\delta F_{\alpha\beta} = \partial_{\alpha}\delta Y_{\beta}
-\partial_{\beta}\delta Y_{\alpha} + [\delta Y_{\alpha},Y_{\beta}]
+ [Y_{\alpha},\delta Y_{\beta}] + [\delta Y_{\alpha},W_{\beta}]+
[W_{\alpha},\delta Y_{\beta}]\,
}
and $F_{\alpha\beta}$ is given by the formula \eqref{Fsplit}.
Setting $\lambda = 0$ we get, 
by \eqref{einsta3}, \eqref{gflatupB}, and the expansions of 
proposition \ref{HeilA}, that  
$D_{2}\Xi^{00}(0,\U_{b},0,0) = 0$.
Therefore $\D_{2}\U^{00}(0,0,0)=0$ by \eqref{redsolA9}. Similar calculations
show that $\D_{3}\U^{00}(0,0,0)=0$.
\end{proof}

%% file: ymd.tex
\sect{ymdsol}{Solving the YMd equations}

Suppose $-1<\delta_{1} < 0$, $-2<\delta_{2}<-1$, $p>3$ and
let
$\Lambda$, $\epsilon$ and $\Uh$ be
as in proposition \ref{redsolA}. Then by the results
of propositions \ref{smoothD} and \ref{redsolA} the
map 
\leqn{YMdC1}{
\Uph : (-\Lambda,\Lambda) \times  B_{\Ac^{2,p}_{\delta_{2}}}(0;\epsilon)\times
B_{\Dc^{2,p}_{\delta_{1}}}(0;\epsilon) \rightarrow \Dc^{0,p}_{\delta_{1}-2}
\times \Ac^{0,p}_{\delta_{2}-2}
}
defined by
\leqn{YMdC2}{
\Uph(\lambda,Y,\xi) := \Upsilon(\lambda,\Uh(\lambda,Y,\xi),Y,\xi)
}
is $C^{\infty}$. 
Define 
\eqn{ChristA}{
\hat{\Gamma}^{\alpha}_{\beta\gamma}(\lambda,Y,\xi) :=
\Gamma^{\alpha}_{\beta\gamma}(\lambda,\Uh(\lambda,Y,\xi)) \, .
}
Then \eqref{Christ}, \eqref{einsta3}, \eqref{gflatupB}, the expansions of 
proposition \ref{HeilA}, and proposition \ref{redsolA} show
that
\eqn{ChristB}{
\hat{\Gamma}^{\alpha}_{\beta\gamma}(0,0,0) = 0\, , \;
\D_{2}\hat{\Gamma}^{\alpha}_{\beta\gamma}(0,0,0) = 0\, \;\text{and}\;
\D_{1}\hat{\Gamma}^{\alpha}_{\beta\gamma}(0,0,0) = 0 \, .
}
Using this result along with \eqref{einsta3}, \eqref{divA2},  \eqref{gflatupB},
and the expansions of proposition \ref{HeilA}, we find after
straightforward calculation that
\lgath{YMdD}{
\Uph^{2}(0,0,0) = \Delta \aep - 
\coup e^{2\kappa\aep}\delta^{ij}\delta^{kl}\ip{F^{\Wep}_{ik}}
{F^{\Wep}_{jl}} \label{YMdD1} \\
\Uph^{1}(0,0,0) = \big( \delta^{ik}\big(\partial_{k}F^{\Wep}_{ij}
+ 2\kappa F^{\Wep}_{ij}\partial_{k}\aep + [\Wep_{k},F^{\Wep}_{jk}]\big)\big) \label{YMdD2}
}
and
\lalign{YMdE}{
\D_{2}\Uph^{1}(0,0,0)\cdot \delta Y & =
\big( \Delta \delta Y_{j} +
 \delta^{ik}\big( [\delta Y_{i}, \partial_{k}\Wep_{j}]  \notag \\
+[\Wep_{i},&\partial_{k}\delta Y_{j}] 
+   2\kappa\Fep(\delta Y)_{ij}\partial_{k}\aep + [\delta Y_{k},F^{\Wep}_{ij}]
+[\Wep_{k},\Fep(\delta Y)_{ij}]
\big) \big) \, , \label{YMdE1} \\
\D_{3}\Uph^{1}(0,0,0)\cdot \delta\xi&=
\big( 2 \kappa\delta^{ik}F^{\Wep}_{ij}\partial_{k} \delta\xi\big) \, , \label{YMdE2} \\
\D_{2}\Uph^{2}(0,0,0)\cdot \delta Y &= 
-2\coup e^{2\kappa\aep}\delta^{ij}\delta^{kl}
\ip{F^{\Wep}_{ik}}{\Fep(\delta Y)_{jl}} \, , \label{YMdE3}\\
\D_{3}\Uph^{2}(0,0,0)\cdot \delta\xi &=
\Delta \delta\xi - 2\kappa\coup \delta\xi
e^{2\kappa\aep}\delta^{ij}\delta^{kl}\ip{F^{\Wep}_{ik}}
{F^{\Wep}_{jl}} \, , \label{YMdE4}
}
where 
\leqn{vF}{
\Fep(\delta Y)_{\alpha\beta} := \partial_{\alpha}\delta Y_{\beta}
-\partial_{\beta}\delta Y_{\alpha} + 
 [\delta Y_{\alpha},\Wep_{\beta}]+ [\Wep_{\alpha},\delta Y_{\beta}]\, .
}
Observe that
\leqn{YMdIII}{
\Uph^{1}(0,0,0) = 0 \quad \text{and} \quad
\Uph^{2}(0,0,0) = 0
}
since $(\Wep,\aep)$ satisfies \eqref{YMdI1}-\eqref{YMdI2}.

We can collect \eqref{YMdE1}-\eqref{YMdE4} into a single matrix expression
\leqn{YMdG}{
\Kcep \begin{pmatrix}\delta Y \\ \delta \xi \end{pmatrix}
= \begin{pmatrix} \Delta & 0 \\ 0 & \Delta \end{pmatrix} 
\begin{pmatrix}\delta Y \\ \delta \xi \end{pmatrix} 
+ K\begin{pmatrix}\delta Y \\ \delta \xi \end{pmatrix}
}
where
\eqn{Kmatrix}{
K := \begin{pmatrix} \Kep_{11} & 
\Kep_{12} \\ \Kep_{21} & \Kep_{22} \end{pmatrix}\, ,}
and
\lalign{YMdH}{
\Kep_{11}\cdot \delta Y& := \big(\delta^{ik}\big( [\delta Y_{i}, \partial_{k}\Wep_{j}]
+[\Wep_{i},\partial_{k}\delta Y_{j}] \notag \\
&+   2\kappa\Fep(\delta Y)_{ij}\partial_{k}\alpha + [\delta Y_{k},F^{\Wep}_{ij}]
+[\Wep_{k},\Fep(\delta Y)_{ij}]
\big) \big) \, , \label{YMdH1} \\
\Kep_{12}\cdot \delta\xi &:=
\big( 2\kappa \delta^{ik}F^{\Wep}_{ij}\partial_{k} \delta\xi\big) \, ,\label{YMdH2} \\
\Kep_{21}\cdot \delta Y &:=
-2\coup e^{2\kappa\aep}\delta^{ij}\delta^{kl}
\ip{F^{\Wep}_{ik}}{\Fep(\delta Y)_{jl}} \, , \label{YMdH3}\\
\Kep_{22}\cdot \delta\xi &:=
- 2\kappa \coup\delta\xi
e^{2\kappa\aep}\delta^{ij}\delta^{kl}\ip{F^{\Wep}_{ik}}
{F^{\Wep}_{jl}} \, . \label{YMdH4}
}

In order to use the implicit function theorem we need
that 
\eqn{iso1}{
\Kcep \; : \;
\Ac^{2,p}_{\delta_{2}}\times \Dc^{2,p}_{\delta_{1}} \longrightarrow
\Ac^{0,p}_{\delta_{2}-2}\times \Dc^{0,p}_{\delta_{1}-2}
}
is an isomorphism. As the next result shows, it will be enough to
establish that $\ker \Kcep = \{0\}$.

\begin{prop} \label{isoA} \mnote{[isoA]}
$\ker \Kcep = \{0\}$ if and only if
 $\Kcep$
is an isomorphism.
\end{prop}
\begin{proof}
Since $-2 < \delta_{2}<-1$  and $-1<\delta_{1} < 0$, 
there exist and $\epsilon > 0$ such that 
$\Kep(\Ac^{2,p}_{\delta_{2}}\times \Dc^{2,p}_{\delta_{1}})
\subset \Ac^{1,p}_{\delta_{2}-(2+\epsilon)}\times \Dc^{1,p}_{\delta_{1}-(1+\epsilon)} $ by
lemma \ref{multiply}.
But the embedding
$ \Ac^{1,p}_{\delta_{2}-(2+\epsilon)}\times \Dc^{1,p}_{\delta_{1}-(2+\epsilon)} 
\rightarrow  \Ac^{0,p}_{\delta_{2}-2}\times \Dc^{0,p}_{\delta_{1}-2}$
is compact by lemma \ref{embed} and hence
$\Kep  :  \Ac^{2,p}_{\delta_{2}}\times \Dc^{2,p}_{\delta_{1}}
\rightarrow  \Ac^{0,p}_{\delta_{2}-2}\times \Dc^{0,p}_{\delta_{1}-2}$
is compact. As
$\Delta \oplus \Delta : \Ac^{2,p}_{\delta_{2}}\times \Dc^{2,p}_{\delta_{1}}
\rightarrow  \Ac^{0,p}_{\delta_{2}-2}\times \Dc^{0,p}_{\delta_{1}-2}$
is an isomorphism by  propositions \ref{slaplaceA} and \ref{slaplaceC}
it follows by compactness of $\Kep$ that
$\text{Index}\left( \Delta \oplus \Delta + \Kep \right) = 0$ 
and the proof is complete. 
\end{proof}

The difficulty
in proving that $\ker{\Kcep} = \{0\}$ lies
with the fact that the spectrum of $\Kcep$ contains both a (strictly) negative
and positive component. The negative
part of the spectrum accounts for
the well known instability of the Yang-Mills-dilaton solutions.
It also means that one cannot expect that $\ker{\Kcep}=\{0\}$ can
be proved by a integration by parts argument. 

\begin{prop} \label{isoB} \mnote{[isoB]}
If \eqref{asymlin2} is valid, then
\eqn{isoB0}{
\Kcep \; :\; \Ac^{2,p}_{\delta_{2}}\times \Dc^{2,p}_{\delta_{1}}
\rightarrow  \Ac^{0,p}_{\delta_{2}-2}\times \Dc^{0,p}_{\delta_{1}-2}
}
is an isomorphism.
\end{prop}
\begin{proof}
Suppose $(\delta Y,\delta\xi) \in \Ac^{2,p}_{\delta_{2}}\times
 \Dc^{2,p}_{\delta_{1}} $ is a solution to 
\leqn{isoB3}{
\Kc \begin{pmatrix} \delta Y\\ \delta \xi 
\end{pmatrix} = 0 \, .
}
We observe that $\Kcep$ is
uniformly elliptic and has $C^{\infty}$ coefficients since
$\Wep$ and $\aep$ are $C^{\infty}$ by \eqref{existA}. Therefore
by elliptic regularity, see \cite{GilTru98} theorem 9.19 or 
\cite{Giaq93} theorem 3.6,
$\delta Y\in \text{C}^{\infty}\cap \Ac^{2,p}_{\delta_{2}}$
and $\delta \xi \in \text{C}^{\infty}\cap  \Dc^{2,p}_{\delta_{1}}$.
Letting 
\eqn{isoB4}{
\phi = \delta \xi \quad \text{and} \quad
\delta Y_{i} = \frac{v(r)}{r^{2}}\epsilon_{i}{}^{j}{}_{k}
x^{k}\tau_{j},
}
shows that $(v(r),\phi(r))$ satisfy the equations 
\eqref{isoB5.1}-\eqref{isoB5.2}. Also since $\delta Y$ and
$\delta \xi$ are smooth it follows that $v(r)$ and $\phi(r)$
satisfy the boundary condition \eqref{bound}. From
our discussion in section \ref{slYMd}, we know that
there must exist constant $c_{i}$ $i=1,2,3,4$ such
that $v(r) = c_{1}v_{1}(r) + c_{2}v_{2}(r)$ and
$\phi(r) = c_{3}\phi_{1}(r) + c_{3}\phi_{2}(r)$.
Assuming that $v_{2}(r)$ satisfies \eqref{asymlin2}
it then follows from \eqref{asymlin1} that 
$\delta \xi \notin \Dc^{2,p}_{\delta_{1}}$ and
$\delta Y \notin \Ac^{2,p}_{\delta_{2}}$ and
hence $\ker \Kcep = \{0\}$.
\end{proof}

We are now ready to solve the YMd equations.

\begin{prop} \label{ymdsolA} \mnote{[ymdsolA]}
Suppose $-1<\delta_{1} < 0$, $-2<\delta_{2} < -1$, $p>3$ and let
$\Lambda$ and $\epsilon$ be as in \eqref{YMdC1}.
If $(w(r),\alpha(r))$ is one of the solutions from theorem \eqref{existA} 
of the Euclidean YMd equations \eqref{YMdI1}-\eqref{YMdI2}
and \eqref{asymlin2} holds 
then there exists $\hat{\Lambda} \in (0,\Lambda)$ and two $\emph{\text{C}}^{\infty}$
maps
\eqn{ymdsoA2}{
\hat{Y} : (-\hat{\Lambda},\hat{\Lambda})  \rightarrow B_{\Ac^{2,p}_{\delta_{2}}}(0;\epsilon) \quad
\text{and} \quad \hat{\xi} : (-\hat{\Lambda},\hat{\Lambda}) \rightarrow
B_{\Dc^{2,p},{\delta_{1}}}(0;\epsilon)
}
such that 
$\hat{Y}(0) = 0$, $\hat{\xi}(0) = 0$
and
$\Uph(\lambda,\hat{Y}(\lambda),\hat{\xi}(\lambda)) = 0$ for all
$\lambda$ $\in (-\hat{\Lambda},\hat{\Lambda})$. 
\end{prop}
\begin{proof}
Because 
$\Kc : \Ac^{2,p}_{\delta_{2}}\times \Dc^{2,p}_{\delta_{1
}}
\rightarrow  \Ac^{0,p}_{\delta_{2}-2}\times \Dc^{0,p}_{\delta_{1}-2}$
is an isomorphism by propositions \ref{isoB} we can apply the implicit functions
theorem to get the desired result.
\end{proof}

%% file: eymdsol.tex
\sect{eymdsol}{Solving the EYMd field equations}

By the propositions \ref{redsolA} and \ref{ymdsolA} we can solve
the reduced field equations \eqref{reduced3} and
the YMd equations \eqref{Ymda1}-\eqref{Ymda2}.
Using the following result of Heilig \cite{Heil95}, we will see that this solution
will actually be a solution to the full EYMd equations.

\begin{prop}{\emph{[proposition 6.1,\cite{Heil95}]}} \label{eymdsolA} \mnote{[eymdsolA]}
Suppose $-1<\delta < 0$, $p>3$, and $\Lambda > 0$. Furthermore, suppose 
\gath{eymdsolA1}{
T : [0,\Lambda] \rightarrow \emph{\W}^{0,p}_{\delta-2}(\Rbb^{3},\Sbb^{3}) \cap
\emph{\text{C}}^{1}(\Rbb^{3},\Sbb^{3}) \; : \; \lambda \mapsto 
(T^{\alpha\beta}_{\lambda} )
\intertext{and}
\U : [0,\Lambda] \rightarrow  \emph{\W}^{2,p}_{\delta}(\Rbb^{3},\Sbb^{3}) \;
: \; \lambda \mapsto (\U^{\alpha\beta}_{\lambda})
}
are two continuous maps such that for every $\lambda \in [0,\Lambda]\;$: 
$(\lambda,\U^{\alpha\beta}_{\lambda},T^{\alpha\beta}_{\lambda})$
is a solution to the reduced field equations \ref{reduced2},
$\nabla_{\beta} T^{\alpha\beta}_{\lambda} = 0$, and
$\partial_{\gamma} T^{\alpha\beta}_{\lambda} 
\in B_{\emph{\W}^{0,p}_{\delta-2}(\Rbb^{3})}(0,R)$ for some $R > 0$ independent
of $\lambda$ and $\alpha,\beta,\gamma$. Then there exists a constant $\hat{\Lambda} \in
(0,\Lambda]$ such that 
$\partial_{\alpha} \U^{\alpha\beta}_{\lambda} = 0$ for all 
$\lambda \in [0,\hat{\Lambda}]$.
\end{prop}

We are now ready to show that to each one of the Euclidean YMd
solutions $(w_{n}(r),\alpha_{n}(r))$ $n=1,2,3,\ldots$
from theorem \ref{existA} for which \eqref{asymlin1}
holds, there exists a solution to the full EYMd
equations. 
\begin{thm} \label{eymdsolB} \mnote{[eymdsolB]}
Suppose $-1 < \delta_{1} < 0$, $-2 < \delta_{2} < 1$, $p>3$ and
let $(w_{n}(r),\phi_{n}(r))$ be one of the solutions 
to the Euclidean Yang-Mills-dilaton equations
from theorem \ref{existA}. If condition \eqref{asymlin2}
holds for the solution $(w_{n}(r),\alpha(r))$ then
there exist a $\Lambda > 0$ and  
$C^{\infty}$ maps
$\U : [-\Lambda, \Lambda] \rightarrow \Uc^{2,p}_{\delta_{2}} \; : \; 
\lambda \mapsto (\U^{\alpha\beta}_{\lambda})\,$ ,  
$Y : [-\Lambda, \Lambda] \rightarrow \Ac^{2,p}_{\delta_{2}} \; : \;
\lambda \mapsto (Y^{\lambda}_{\alpha}) \,$ ,
and $\xi : [-\Lambda, \Lambda] \rightarrow \Dc^{2,p}_{\delta_{1}} \; : \;
\lambda \mapsto \xi^{\lambda}$ 
such that $(Y^{0},\xi^{0}) = (0,0)$ and for any $\lambda \in (0,\Lambda]\;$ 
$(\lambda,\U^{\alpha\beta}_{\lambda},A^{\lambda} = W^{n} + Y^{\lambda},\psi^{\lambda} = \alpha_{n} +
\xi^{\lambda})$ is
a $\emph{\text{C}}^{2}$ solution to the EYMd 
equations \eqref{einst}, \eqref{Ymda1}, and \eqref{Ymda2}. Moreover,
the solution is static, spherically symmetric, and asymptotically
flat.
\end{thm}
\begin{proof}
Let $(w_{n},\alpha_{n})$ be one of the solutions
 to the Euclidean YMd
equations from theorem \ref{existA} and let
$W^{n}_{\alpha} = \delta_{\alpha}^{i}
r^{-2}(w_{n}(r)-1)\epsilon_{i}{}^{j}{}_{k}x^{k}\tau_{j}$. 
If we assume that \eqref{asymlin2} holds for the
solution $(w_{n}(r),\alpha_{n}(r)$ then by
propositions 
\ref{redsolA} and \ref{ymdsolA} 
there exists a $\Lambda > 0$ and $\text{C}^{\infty}$ maps
$\U : [-\Lambda, \Lambda] \rightarrow \Uc^{2,p}_{\delta_{2}}$,
$\; :\;\lambda \mapsto (\U^{\alpha\beta}_{\lambda})$,
$Y : [-\Lambda, \Lambda] \rightarrow \Ac^{2,p}_{\delta_{2}}$ $\; : \;
\lambda \mapsto (Y^{\lambda}_{\alpha})$ , 
and
$\xi : [-\Lambda, \Lambda] \rightarrow \Dc^{2,p}_{\delta_{1}}$$ \; : \;
\lambda \mapsto \xi^{\lambda}$,
such that $(Y^{0},\xi^{0}) = (0,0)$, and 
\eqn{eymdsolB3a}{
\Xi(\lambda,\U(\lambda),Y(\lambda),\xi(\lambda)) = 0\, , \quad
\Upsilon(\lambda,\U(\lambda),Y(\lambda),\xi(\lambda)) = 0 
}
for all $\lambda \in (-\Lambda,\Lambda) $. 
To reduce notation, we will often write $\U$, $Y$ and $\xi$ instead of 
$\U_{\lambda}$, $Y^{\lambda}$,
and $\xi^{\lambda}$. 

\begin{lem} \label{eymdsolC} \mnote{[eymdsolC]}
There exists a $\Lambda^{*} \in (0,\Lambda]$ such that
$A^{\lambda} = W^{n} + Y^{\lambda}, \psi^{\lambda} = \alpha_{n} + \xi^{\lambda} \in \emph{\text{C}}^{2}$ for all
$\lambda \in (-\Lambda^{*},\Lambda^{*})$.
\end{lem}
\begin{proof}
Let $\text{B}_{R} \subset \Rbb^{3}$ be an open ball of radius R centered at the origin.
Then $\psi$, $\psi_{,\alpha}$, $\U^{\alpha\beta}$, $\U^{\alpha\beta}{}_{,\mu}$
, $A_{\alpha}$, $A_{\alpha,\beta}$ 
$\in \W^{1,p}(\text{B}_{R})$, where recall that $A = W^{n} + Y$ 
and $\psi = \alpha_{n} + \xi$. 
As $\W^{1,p}(\text{B}_{R})$
is a Banach algebra, we have
\gath{eymdsolB3}{
f := \Gamma^{\mu}_{\alpha\beta}\psi_{,\mu}\gb^{\alpha\beta} - \frac{\kappa\ell_{Y}}{\ell_{d}}
\gb^{\alpha\mu}\ip{F_{\alpha\mu}}{F_{\alpha\beta}} \in \W^{1,p}(\text{B}_{R}) \,
\\
h = (h_{j}) := \big(\gb^{\alpha\nu}\left( \Gamma^{\mu}_{\alpha \nu} F_{\mu j} +
\Gamma^{\mu}_{j\nu} F_{\alpha\mu} - 2\kappa \psi_{,\nu} F_{\alpha j} - [A_{\nu},
F_{\alpha,j}]\right) \big) \in \W^{1,p}(\text{B}_{R},\Rbb^{3})\, ,
\\
\gb^{ij} = \delta^{ij} + 4\lambda^{2}\U^{ij} \in W^{1,p}(\text{B}_{R}) \, 
\\
Q^{ik} = \left(Q^{ikl}{}_{j}\right) := \big( (\delta^{ik}+4\lambda^{2}\U^{ik})
\delta^{l}_{j} - 4\lambda^{2}\U^{lk}\delta^{i}_{j}\big) \in \W^{1,p}(\text{B}_{R},
\mathbb{M}_{3\times 3})
}
and hence
$f, h_{j}, \gb^{ij}, Q^{ikl}{}_{j} \in \text{C}^{0,1-3/p}(\text{B}_{R})$
by the Sobolev embedding theorem. Notice that YMd equations
$\Upsilon(\lambda,\U(\lambda),Y(\lambda),\xi(\lambda)) = 0$ can be written as
\eqn{eymdsolB5}{
\gb^{ij}\partial^{2}_{ij}\psi = f 
\quad \text{and} \quad Q^{ijl}{}_{j}\partial^{2}_{ij}A_{l} = h_{j} \, . 
}
By the weighted Sobolev inequality, \cite{Bart86} theorem 1.2 (v), the
embedding $\W^{1,p}_{\delta_{1}}(\Rbb^{3},\Sbb^{3}) 
\rightarrow C^{0,1-3/p}_{\delta_{1}
}(\Rbb^{3},\Sbb^{3})$ is
continuous and hence the map $(-\Lambda,\Lambda) \rightarrow
C^{0,1-3/p}_{\delta_{1}}
(\Rbb^{3},\Sbb^{3}) \: : \: \lambda \mapsto U(\lambda)$ is
continuous.
Therefore there exists a $\Lambda^{*} \in (0,\Lambda)$ such that
the operators
$\gb^{ij}\partial^{2}_{ij}$ and $Q^{ij}\partial^{2}_{ij}$
are uniformly elliptic with coefficients in $C^{0,1-3/p}_{\delta_{1}
}(\Rbb^{3})$
 for all $\lambda \in [-\Lambda^{*},\Lambda^{*}]$. By
elliptic regularity, $A^{\lambda}_{j} = W_{j} + Y^{\lambda}_{j}$ and $\psi^{\lambda}
 = \alpha + \psi^{\lambda}$ are
in $\text{C}^{2}(\text{B}_{R})$
for all $\lambda  \in [-\Lambda^{*},\Lambda^{*}]$. As $\Lambda^{*}$ is independent of
$R$ the result follows.
\end{proof}

It follows immediately from equation \eqref{streng1a}, proposition \ref{smoothB},
and the above lemma that the hypotheses of proposition \ref{eymdsolA} are satisfied.
Therefore we conclude that there exist a constant $\hat{\Lambda} \in (0,\Lambda^{*}]$ such
that  
\leqn{eymdsolB7}{
\partial_{\alpha}\U^{\alpha\beta}_{\lambda} = 0
}
for all $\lambda \in [0,\hat{\Lambda}]$.
This implies that the full EYMd equations are equivalent to
$\Xi = 0$ and $\Upsilon = 0$ and hence 
$(\lambda,\U_{\lambda},A^{\lambda}=W+Y^{\lambda},\psi^{\lambda}=\alpha + \xi^{\lambda})$
satisfy the EYMd equations for all $\lambda \in (0,\hat{\Lambda}]$.

Using \eqref{eymdsolB7}, the reduced field equations $\Xi=0$ can be written as
\eqn{eymdsolB8}{
\gb^{ij}\partial^{2}_{ij}\U^{\alpha\beta} = H^{\alpha\beta} 
}
where $H^{\alpha\beta} = 
-A^{\alpha\beta} - B^{\alpha\beta} - 
C^{\alpha\beta} + 4\pi G |\df| T^{\alpha\beta}$. 
We can then argue as in lemma \ref{eymdsolC} to
conclude that $\U^{\alpha\beta} \in \text{C}^{2}$.
\end{proof}

%% file: conc.tex
\sect{conc}{Conclusion}

In this paper we have shown how to reduce the existence
problem for static spherically symmetric solutions
to the $SU(2)$ EYMd equations to that of proving
the non-existence of solutions to the lYMd equations
by using a Newtonian perturbation argument. We 
conjectured that solutions to the lYMd
equations satisfy \eqref{asymlin2} and showed that if
this is true 
then there exists a countably infinite
number of static spherically symmetric solutions.

Numerically it has been
found that the EYMd equations also admit static axially
symmetric solutions \cite{KK98}. There is nothing
in principle from generalizing the results of this
paper to the non-spherically symmetric case. 
To make progress in the non-spherically
symmetric case a PDE proof that the lYMd equations
have only the trivial solution would be needed. However,
as discussed in this paper, even in the
spherically symmetric case, this is a difficult
problem and would represent a significant 
advance. The other main problem would
be to try and prove that static axially symmetric
solutions to the YMd equations exist. Even though
this problem would be much simpler than proving
the existence of solutions to the full EYMd equations
it still represents an extremely difficult problem.